\input jytex.tex   
\typesize=10pt
\magnification=1200
\baselineskip17truept
\hsize=6truein\vsize=8.5truein
\sectionnumstyle{blank}
\chapternumstyle{blank}
\chapternum=1
\sectionnum=1
\pagenum=0

\def\begintitle{\pagenumstyle{blank}\parindent=0pt\begin{narrow}[0.4in]}
\def\endtitle{\end{narrow}\newpage\pagenumstyle{arabic}}


\def\beginexercise{\vskip 20truept\parindent=0pt\begin{narrow}[10 
truept]}
\def\endexercise{\vskip 10truept\end{narrow}}


\def\eql#1{\eqno\eqnlabel{#1}}
\def\ref{\reference}
\def\peq{\puteqn}
\def\pref{\putref}

\def\mgn{\marginnote}
\def\bex{\begin{exercise}}
\def\eex{\end{exercise}}


\font\open=msbm10 
\def\mbox#1{{\leavevmode\hbox{#1}}}

\def\hspace#1{{\phantom{\mbox#1}}}
\def\oZ{\mbox{\open\char90}}

\def\rS{{\rm S}}

\def\al{\alpha}
\def\bka{{\bmit\kappa}}
\def\bom{{\bmit\omega}}
\def\be{\beta}
\def\ga{\gamma}
\def\de{\delta}
\def\Ga{\Gamma}

\def\La{\Lambda}
\def\om{\omega}

\def\si{\sigma}
\def\Si{\Sigma}

\def\ze{\zeta}

\def\De{\Delta}

\def\ball{{\rm ball}}

\def\Res{{\rm Res\,}}
\def\zf{$\zeta$--function}
\def\zfs{$\zeta$--functions}
\def\noa{\noalign{\vskip 5truept}}

\def\frac#1/#2{\leavevmode\kern.1em
\raise.5ex\hbox{\the\scriptfont0 #1}\kern-.1em/\kern-.15em
\lower.25ex\hbox{\the\scriptfont0 #2}}
\def\sfrac#1/#2{\leavevmode\kern.1em
\raise.5ex\hbox{\the\scriptscriptfont0 #1}\kern-.1em/\kern-.15em
\lower.25ex\hbox{\the\scriptscriptfont0 #2}}

\def\gtorder{\mathrel{\raise.3ex\hbox{$>$}\mkern-14mu
             \lower0.6ex\hbox{$\sim$}}}
\def\ltorder{\mathrel{\raise.3ex\hbox{$<$}\mkern-14mu
             \lower0.6ex\hbox{$\sim$}}}

\def\semidirprod{\rlap{\ss C}\raise1pt\hbox{$\mkern.75mu\times$}}
\def\for{\lower6pt\hbox{$\Big|$}}
\def\fish{\kern-.25em{\phantom{abcde}\over \phantom{abcde}}\kern-.25em}


\def\boxit#1{\vbox{\hrule\hbox{\vrule\kern3pt
        \vbox{\kern3pt#1\kern3pt}\kern3pt\vrule}\hrule}}
\def\dalemb#1#2{{\vbox{\hrule height .#2pt
        \hbox{\vrule width.#2pt height#1pt \kern#1pt
                \vrule width.#2pt}
        \hrule height.#2pt}}}

\def\frac#1#2{{{#1}\over{#2}}}

\def\noin{\noindent}

\def\comb#1#2{{\left(#1\atop#2\right)}}

\def\eg{{\it e.g. }}
\def\ie{{\it i.e. }}
\def\cf{{\it cf }}
\def\pa{\partial}


\def\tr{{\rm tr\,}}

\def\3j#1#2#3#4#5#6{\left\lgroup\matrix{#1&#2&#3\cr#4&#5&#6\cr}
\right\rgroup}

\def\caz{{\cal Z}}
\def\man{{\cal M}}
\def\can{{\cal N}}

\def\m?{\mgn{?}}

\def\pa{\partial}

\def\beq{\begin{eqnarray}}
\def\eeq{\end{eqnarray}}
\def\zb{\zeta_{{\cal B}}(}

\def\zn{\zeta_{{\cal N}}\left(} 
\def\rzn{\Res\,\,\zn}


\def\aop#1#2#3{{\it Ann. Phys.} {\bf {#1}} (19{#2}) #3}

\def\cmp#1#2#3{{\it Comm. Math. Phys.} {\bf {#1}} (19{#2}) #3}
\def\cqg#1#2#3{{\it Class. Quant. Grav.} {\bf {#1}} (19{#2}) #3}

\def\jgp#1#2#3{{\it J. Geom. and Phys.} {\bf {#1}} (19{#2}) #3}
\def\jmp#1#2#3{{\it J. Math. Phys.} {\bf {#1}} (19{#2}) #3}
\def\jpa#1#2#3{{\it J. Phys.} {\bf A{#1}} (19{#2}) #3}

\def\np#1#2#3{{\it Nucl. Phys.} {\bf B{#1}} (19{#2}) #3}
\def\pl#1#2#3{{\it Phys. Lett.} {\bf {#1}} (19{#2}) #3}

\def\prD#1#2#3{{\it Phys. Rev.} {\bf D{#1}} (19{#2}) #3}

\def\cras#1#2#3{{\it Comptes Rend. Acad. Sci. (Paris)} {\bf{#1}} (#2) #3}

\def\mpcps#1#2#3{{\it Math. Proc. Camb. Phil. Soc.} {\bf{#1}} (19{#2}) #3}

\def\am#1#2#3{{\it Acta Mathematica} {\bf {#1}} (19{#2}) #3}
\def\aim#1#2#3{{\it Adv. in Math.} {\bf {#1}} (19{#2}) #3}
\def\ajm#1#2#3{{\it Am. J. Math.} {\bf {#1}} ({#2}) #3}

\def\aom#1#2#3{{\it Ann. of Math.} {\bf {#1}} (19{#2}) #3}

\def\cpde#1#2#3{{\it Comm. Partial Diff. Equns.} {\bf {#1}} (19{#2}) #3}

\def\invm#1#2#3{{\it Invent. Math.} {\bf {#1}} (19{#2}) #3}
\def\ijpam#1#2#3{{\it Ind. J. Pure and Appl. Math.} {\bf {#1}} (19{#2}) #3}
\def\jdg#1#2#3{{\it J. Diff. Geom.} {\bf {#1}} (19{#2}) #3}

\def\jmpa#1#2#3{{\it J. Math. Pures. Appl.} {\bf {#1}} ({#2}) #3}

\def\ojm#1#2#3{{\it Osaka J.Math.} {\bf {#1}} ({#2}) #3}

\def\pja#1#2#3{{\it Proc. Jap. Acad.} {\bf {A#1}} (19{#2}) #3}

\def\tams#1#2#3{{\it Trans. Am. Math. Soc.} {\bf {#1}} (19{#2}) #3}

\begin{title}  
\vglue 1truein
\righttext {MUTP/96/23}
\righttext{hep-th/96}
\vskip15truept
\centertext {\Bigfonts \bf Spinors and forms on generalised cones}
\vskip 20truept 
\centertext{J.S.Dowker\footnote{dowker@a3.ph.man.ac.uk},\quad
Klaus Kirsten\footnote{kirsten@tph100.physik.uni-leipzig.de}} 
\vskip 7truept
\centertext{*\it Department of Theoretical Physics,\\
The University of Manchester, Manchester, England}
\vskip10truept
\vskip10truept
\vskip7truept
\centertext{\dag\it Universit\"at Leipzig, Institut f\"ur Theoretische 
Physik,\\ Augustusplatz 10, 04109 Leipzig, Germany}
\vskip 20truept
\centertext {Abstract}
\vskip10truept
\begin{narrow}
A method is presented, and used, for determining any heat-kernel
coefficient for the form-valued Laplacian on the $D$-ball as an explicit 
function of dimension and form order. The calculation is offerred as a
particular application of a general technique developed earlier 
for obtaining heat-kernel
coefficients on a bounded generalised cone which involves writing the 
sphere and ball \zfs, and coefficients, in terms of Barnes \zfs\ and
generalised Bernoulli polynomials. Functional determinants are  
computed. Spinors are also treated by the general method.
\end{narrow}
\vskip 5truept
\righttext {August 1996}
\vskip 60truept
\vfil
\end{title}
\pagenum=0
\section {\bf 1. Introduction}

The present work should be considered as an extension of an earlier one,
[\pref{BKD}] where heat-kernel asymptotics and functional 
determinants were computed 
for a scalar field on a bounded generalised cone (the simplest example of 
which is a ball). Here, the scalar field is replaced by more
complicated bundles, specifically by spinors and forms. This will enable
us later to discuss topological matters such as the analytic torsion
as well as settling a few questions regarding the spectral
invariants. However the emphasis here will be on the actual computations 
and the technical problems raised have, we think, elegant outcomes. 
A primary aim is to work in arbitrary dimensions with arbitrary 
$p$-forms. In this connection, one objective is to make contact with the 
work of Bla\v zi\'k {\it et al} [\pref{BBG}] who have given general, explicit
expressions for the first four heat-kernel coefficients for the form-valued
Laplacian on manifolds with boundary. We anticipate going beyond their
computations but, of course, within the confines of a restricted
geometry. 

The structure of this paper is as follows. In section 2, 
we discuss the various types of eigenforms that exist on the generalised 
cone. For rapidity, we use the results of Cheeger [\pref{cheeg1}] on the 
infinite cone, additionally imposing appropriate conditions at the cone 
boundary. 

Section 3 contains the general construction of the \zf\ and heat-kernel 
coefficients. This closely follows our earlier development so that 
explanatory
details can be kept to a minimum. The result is a formula for the $p$-form 
coefficients on the generalised cone in terms of those on its base
for a particular, `modified' coexact $p$-form.

In sections 4 and 5 this apparatus is applied to the case of the ball 
where it is possible to find a compact expression for the relevant 
$p$-form \zf\ on the sphere in terms of Barnes \zfs. 
Another more suggestive form of this sphere \zf\ expresses it as a sum of 
0--form \zfs. The ball coefficients are determined as explicit 
functions of $d$ and $p$ via generalised Bernoulli functions. 

Rearranging the first 
few coefficients we obtain agreement with the general formulae of 
[\pref{BBG}] evaluated on the ball. Going beyond this, we are able, in 
section 6, to find the expression for an arbitrary coefficient in terms of  
binomial coefficients and polynomials in the dimension. This is a natural 
generalisation of our previous results [\pref{BKD}], and of Levitin's 
earlier [\pref{Levitin}], for the scalar case. Section 7 briefly looks at 
the important value of the \zf\ evaluated at zero argument.

In section 8 we address the calculation of the 
functional determinants  and some explicit results are exhibited in an
appendix.

Related work has been performed by Elizalde {\it et al} [\pref{ELV2}] 
and we use their
specific results as a check of our general forms. Their method of 
evaluation also works generally but involves an immediate reduction to a 
series of Hurwitz \zfs\ and does not yield the functional form. The 
expressions for the functional determinants obtained in [\pref{ELV2}] contain
integrals over $\Psi$ and $\Ga$ functions. In an appendix we show how these
may be reduced to derivatives of the Riemann \zf.

In section 9 we turn, rather more briefly, to spinors and apply our
general method to rapidly reproduce and justify our earlier results for
spectral boundary conditions. A curious vanishing theorem, noticed earlier, 
is proved in an appendix. Local conditions are also treated by the 
improved technique.

\section{\bf 2. Geometry and eigenforms.}

The geometry has been outlined in Cheeger [\pref{cheeg1}] and already 
used in our
earlier work, [\pref{BKD}], so that it will not be described at length
here. The cone manifold, $\man$ has the hyperspherical polar metric
$ds^2=dr^2+r^2d\Si^2$ with $d\Si^2$ the metric on the base, $\can$, 
which is 
the $r=1$ section of the cone. For simplicity, we shall assume in this 
paper 
that the base is closed, $\pa\can=\emptyset$, which is enough for the ball.

The structure of the eigenforms of the de Rham Laplacian, $\De_\man$, 
on the infinite generalised cone has been given by Cheeger [\pref{cheeg1}] 
and, for convenience, we refer to [\pref{cheeg1}] for the detailed action 
of $\De_\man$ on forms of the type $\be+dr\wedge\om$ as well as for the 
necessary exterior calculus. (See also [\pref{cheeg2}].) 

There are four basic types of eigenforms with nonzero eigenvalues, $\al^2$,
$$
\phi^{\man(1)}_p={J_{\nu(p)}(\al r)\over r^{(d-1-2p)/2}}\,\phi^\can_p\,,
\eql{eigform1}
$$

$$
\phi^{\man(2)}_p={J_{\nu(p-1)}(\al r)\over r^{(d+1-2p)/2}}\,
\tilde d\phi^\can_{p-1}+\bigg({J_{\nu(p-1)}(\al r)\over r^{(d+1-2p)/2}}
\bigg)'\,dr\wedge \phi^\can_{p-1},
\eql{eigform2}$$

$$\eqalign{
\phi^{\man(3)}_p={1\over r^{d-2p}}\big(r^{(d+1-2p)/2}\,&J_{\nu(p-1)}
(\al r)\big)'\tilde d\phi^\can_{p-1}\cr+
&{J_{\nu(p-1)}(\al r)\over r^{(d+3-2p)/2}}dr\wedge
\tilde\de\tilde d\phi^\can_{p-1},\cr}
\eql{eigform3}$$

$$
\phi^{\man(4)}_p={J_{\nu(p-2)}(\al r)\over r^{(d+1-2p)/2}}\,dr\wedge
\tilde d\phi^\can_{p-2},
\eql{eigform4}$$
where $\phi^\can_p$ is a coexact $p$-eigenform of the intrinsic 
de Rham Laplacian, $\De_\can=\tilde d\tilde\de+\tilde\de\tilde d$, 
on the base $\can$. We have the separation of variables relation
$$
\nu(p)=\big(\mu(p)+((d-1)/2-p)^2\big)^{1/2}
\eql{reln}$$ with $\mu(p)$ being the (coexact) eigenvalue of $\De_\can$.
We are assuming that $\nu\ge1$ so that the `negative' modes 
$\sim J_{-\nu}$ do not arise. 

In addition, there are modes (`zero modes') whose $\can$ part is harmonic, 
$\mu=0$,
$$
\phi^{\man(E)}_p={J_{\nu_E(p)}(\al r)\over r^{(d-1-2p)/2}}\,h^\can_p
\eql{eigformE}
$$
and
$$
\phi^{\man(O)}_p=\bigg({J_{\nu_O(p)}(\al r)\over r^{(d+1-2p)/2}}
\bigg)'\,dr\wedge h^\can_{p-1},
\eql{eigformO}$$
with
$$\eqalign{
\nu_E(p)&=|(d-1)/2-p|=\nu_E(d-1-p)\cr
\nu_O(p)&=|(d+1)/2-p|=\nu_E(d-p).\cr}
\eql{znus}$$

On $\man$, types 1, $E$ and 3 are coexact and types 2, $O$ and 4 are 
exact. (See the relations in [\pref{cheeg2}].)

For the bounded, generalised cone, conditions are to be set at $r=1$.
Absolute boundary conditions are [\pref{cheeg3,Gilkey1}]
$$
\big(\phi^{\man}_{i\ldots j}\big)'\bigg|_\can=0,\quad
\phi^\man_{r,i\ldots j}\bigg|_\can=0
\eql{absbc}$$
and have to be applied to the six types separately.

Since $\phi^\can_p$, $h^\can_p$ are pure $\can$ forms, it is easily shown 
that
types 1, 2, $E$ and $O$ satisfy Neumann (Robin) and types 3 and 4 Dirichlet 
conditions. Bessel's equation has to be used to derive this for type 3.

More precisely the Robin conditions are (\eg for type 1)
$$
\pa_r\big(r^{p-(d-1)/2}J_{\nu(p)}(\al r)\big)\bigg|_\can=0
\eql{arc}$$
so that the Robin parameter is $u=u_a(p)=p-(d-1)/2$, 
([\pref{BGKE}] 3.1). The parameter for type 2 is $u_a(p-1)$.

Relative boundary conditions are obtained by dualising and, in the present
context amount to
$$
\big(\pa_r+d+2-2p\big)\phi^{\man}_{r,i\ldots j}\bigg|_\can=0,
\quad\phi^\man_{i\ldots j}\bigg|_\can=0.
\eql{relbc}$$
Now types 1, 2, $E$ and $O$ satisfy Dirichlet and types 3 and 4 Robin 
conditions. For type 3, Robin conditions read
$$
\pa_r\big(r^{(d+1)/2-p}J_{\nu(p-1)}(\al r)\big)\bigg|_\can=0
\eql{rrc}$$
with the Robin parameter $u=u_r(p)=(d+1)/2-p=u_a(d-p)$. The parameter for
type 4 is $u_r(p-1)$. Hodge $\star$-duality on $\man$ thus interchanges 
the conditions (\peq{arc}) and (\peq{rrc}) with $1\leftrightarrow4$ 
and $2\leftrightarrow3$ which differs from Cheeger.

We denote the coexact degeneracies on $\can$ by $d(p)$ and remark that the 
exact degeneracies, $d_{ex}(p)$, and eigenvalues, $\mu_{ex}(p)$, are given 
by
$$
d_{ex}(p)=d(p-1),\quad
\mu_{ex}(p)=\mu(p-1).
\eql{degrel}$$
The structure of the eigenforms shows that the degeneracies of types 
1, 2, 3 and 4 are $d(p)$, $d(p-1)$, $d(p-1)$ and $d(p-2)$ respectively.

We now define, for later use, the modified coexact \zf\ on $\can$, 
the base \zf,
$$
\ze^\can_p(s)=\sum {d(p)\over\nu(p)^{2s}}=\sum {d(p)
\over\big(\mu(p)+((d-1)/2-p)^2\big)^s}
\eql{modcoexzet}$$
(\cf [\pref{BKD}] (3.3).) Including the `zero modes', which is sometimes
notationally convenient, one has
$$
\tilde\ze^\can_p(s)=\ze^\can_p(s)+{\be^\can_p\over\nu_E(p)^{2s}}
\equiv\sum {\tilde d(p)\over\nu(p)^{2s}},
\eql{zetzeroin}$$
which defines $\tilde d(p)$ and where $\be_p^\can$ is the $p$-th Betti 
number of $\can$.
The summations, here and later, are over the mode
labels which are not always explicitly displayed.

In view of the relations (\peq{degrel}), an `exact'
\zf\ is defined by
$$
\ze^\can_{ex,\,p}(s)=\ze^\can_{p-1}(s).
\eql{modexzet}$$
Because the additional term $\big((d-1)/2-p\big)^2$ depends on $p$,
$\ze^\can_{ex,\,p}(s)$ is not the true exact \zf, but we expect it to be
the relevant construct. 

Assume for the time being that the base $\can$ has no boundary. Hodge
duality on $\can$ can be applied to yield the coexact relations
$$
d(d-1-p)=d(p),\quad{\rm and}\quad \mu(d-1-p)=\mu(p)
\eql{dual1}$$
whence
$$
\nu(d-1-p)=\nu(p)
\eql{dual2}$$
and therefore, for the coexact \zf,
$$
\ze^\can_{d-1-p}(s)=\ze^\can_p(s).
\eql{zedual}$$

These relations will be verified later when $\can$ is a sphere.

\section{\bf 3. General constructions.}
The {\it total} \zf\ on $\man$ is, according to 
(\peq{eigform1}) -- (\peq{eigform4}), (\peq{eigformE}), (\peq{eigformO}), 
a combination of exact and coexact (on $\man$) contributions.
Noting that the zero modes on $\can$ give degenerate eigenvalues on $\man$,
one has
$$
\ze^{\man+}_p(s)=\sum_{i=1}^4\sum_{\al_i}{1\over\al_i^{2s}}
+\be^\can_p\sum_{\al_E}{1\over\al_E^{2s}}
+\be^\can_{p-1}\sum_{\al_O}{1\over\al_O^{2s}}
\eql{totzetm}$$
where the $\al_i^{2s}$ are the eigenvalues of the $p$-form Laplacian on $\man$
for $i$-type modes.

(\peq{totzetm}) can be written in terms of the coexact \zf\ on $\man$,
$\ze^\man_p(s)$, as
$$
\ze^{\man+}_p(s)=\ze^{\man}_p(s)+\ze^{\man}_{p-1}(s),
\eql{totzet2}$$
the inverse of which is
$$
\ze_p^\man(s)=\sum_{q=0}^p(-1)^{p-q}\ze_p^{\man+}(s).
\eql{inver}$$

According to the method given in [\pref{BEK,BGKE,BKD}],
and now for absolute conditions,
$$
\ze^{\man}_{a,p}(s)=\!\!\sum\!
\int_\ga{dk\over2\pi i}k^{-2s}{\pa\over\pa k}\!
\bigg(\tilde d(p)\ln\big(r^{u_a(p)} J_{\nu(p)}(kr)\big)'\bigg|_\can
+d(p-1)\ln J_{\nu(p-1)}(k)\bigg), 
\eql{coexze}$$
where the details of the contour $\ga$ are in the above cited references.
The first term is the Neumann (Robin) (types 1 and $E$) 
and the second term the Dirichlet (type 3) part. 
For economy of writing, the degeneracy $\tilde d(p)$ is introduced so as 
to take into account the $\be^\can_p$ $E$-type zero modes on $\can$, 
\cf (\peq{zetzeroin}). When $p\to p-1$, the type 1 contribution becomes 
a type 2,  the type $E$ a type $O$ and the type 3 a type 4.

For relative conditions, likewise,
$$
\ze^{\man}_{r,p}(s)=\!\!\sum\!
\int_\ga{dk\over2\pi i}k^{-2s}{\pa\over\pa k}
\bigg(d(p-1)\ln\big(r^{u_r(p)}J_{\nu(p-1)}(kr)\big)'\bigg|_\can
+\tilde d(p)\ln J_{\nu(p)}(k)\bigg). 
\eql{rcoexze}$$

It is amusing to check in detail Hodge duality on $\man$, 
$$
\ze^{\man+}_{a,p}(s)=\ze^{\man+}_{r,d+1-p}(s),
\eql{hdual}$$
which fundamentally arises from the intertwining [\pref{Gilkey1}]
$$
\star\,\De^\man_{p,a}=\De^\man_{d+1-p,r}\star.
$$

It is easily seen from (\peq{totzet2}) that (\peq{hdual}) is equivalent
to the statement that, under $p\to d-p$, $\ze^{\man}_{a,p}(s)$ of
(\peq{coexze}) turns into $\ze^{\man}_{r,p}(s)$ of (\peq{rcoexze}).
Firstly, this is readily seen to be the case from the relations 
(\peq{dual1}) and (\peq{dual2}), if the zero $E$ modes are set aside.  
Secondly, consider the contribution coming from the 
zero $E$ modes in (\peq{coexze}). One has $u_a (p) = p-(d-1)/2$ and
the argument of the logarithm reads explicitly
$
u_a (p) J_{\nu_E(p)} +kJ'_{\nu_E(p)} 
$
where $|u_a (p)| = \nu_E(p)$. Let us assume for the moment that
$u_a(p)\leq 0$, and then use the relation
$
z J_{\nu}' (z) -\nu J_{\nu} (z) = - z J_{\nu+1} (z) 
$
to write the above argument as
$
-k J_{\nu_E(p) +1}.
$

The factor $(-k)$ does not contribute because it has no zero
inside the contour, $\gamma$, and so the contribution 
is just that of $J_{\nu_E(p) +1}$.
Now from (\peq{znus}), $ \nu_E (p) +1=\nu_E(d-p)$ 
and, taking into account Poincar\'e duality on $\can$, $\be_p^\can
=\be_{d-p}^\can$, the contribution of the $p$-form
zero modes for absolute boundary conditions is seen to equal
that of the $(d-p)$-form zero modes for relative 
boundary conditions, which was to be shown in order to complete the 
demonstration of Hodge duality on $\man$.
For $u_a (p) > 0$ use
$
z J_{\nu} ' (z) +\nu J_{\nu } (z) = z J_{\nu -1} (z)  
$
to arrive at the same conclusion.

One of our main objectives is the evaluation of the
coefficients, $A_{n/2}$, in the heat-kernel expansion, which is stated in 
generic form,
$$
K^\man(\tau)=\sum_{n=0}^\infty A_{n/2}^\man\,\tau^{(n-D)/2}+A'_\man
\log\tau.
\eql{expans}$$

Our preferred general computational formula, 
on any manifold, $\man$, of dimension $D$, is
$$
A^\man_{n/2}=\Gamma\big((D-n)/2\big)\Res\ze^\man\big((D-n)/2\big),
\quad n<D.
\eql{coefff}$$

The ability, by choosing $D$ sufficiently large, to work with just this 
formula in order to determine {\it any} coefficient, has an additional 
advantage because a finite number of `extra' modes does not affect the 
analytic structure, as in the difference between $\ze$ and $\tilde\ze$ when
(\peq{coefff}) is applied to the base $\can$. 

The calculation of the \zf\ on $\man$ in its relation to that on $\can$, 
follows precisely the previous pattern, [\pref{BKD}], the Robin and 
Dirichlet cases now being combined. It leads
to the following basic equation for the coexact heat-kernel coefficients 
on $\man$ in terms of those on $\can$, ($n<d$),
$$\eqalign{
A^{\man}_{a,n/2}(p)=&{1\over2\sqrt\pi(D-n)}\big({A}^{\can}_{n/2}(p)
+A^{\can}_{n/2}(p-1)\big)\cr
&+{1\over4}\big(A^{\can}_{(n-1)/2}(p)-
A^{\can}_{(n-1)/2}(p-1)\big)\cr
&-\sum_{i=1}^{n-1}\bigg(A^{\can}_{(n-1-i)/2}(p)\,P_{i}\big(z_a(p)\big)
+A^{\can}_{(n-1-i)/2}(p-1)\,P_i(x)\bigg)\cr}
\eql{comancoeff}$$
for absolute conditions.

The $A^\can$ are the heat-kernel coefficients corresponding to the 
base \zf, (\peq{zetzeroin}), and the $P_i$ are known polynomials arising 
from the asymptotic expansion of Bessel functions, \eg
$$
P_i(x)=\sum_{b=0}^i x_{i,b}\, 
{\Gamma\left((D-n+i)/2 +b\right)\over
\Gamma \left((D-n+i)/2\right)\Gamma(b+i/2)}.
\eql{polys}$$

The $x$'s correspond to Dirichlet and the $z_a(p)$'s, which depend on $p$
through the $u_a(p)$, to Robin conditions.

The combination (\peq{totzet2}) leads to the total coefficients on $\man$,
$$
A^{\man+}_{a,n/2}(p)=A^{\man}_{a,n/2}(p)+A^{\man}_{a,n/2}(p-1)
\eql{comb}$$
and similarly for relative conditions. (\peq{comb}) can be inverted as in
(\peq{inver}).

In general $\ze^\man_p(s)$ has a pole at $s=0$ which translates into the 
$\log\tau$ term in the expansion (\peq{expans}), the coefficient being
$$
A'_\man(p)=-\Res\ze^\man_p(0).
\eql{logcoeff}$$

These are the only general equations that are needed but the algebra can be 
checked by confirming Hodge duality on $\man$, in the coefficient form 
$$
A^{\man+}_{r,n/2}(d+1-p)=A^{\man+}_{a,n/2}(p),
\eql{coeffhd}$$
using the easily verified formula,
$$
A^{\can\pm}_{n/2}(d-p)=\pm A^{\can\pm}_{n/2}(p)
$$
where, the coefficients $A^{\can\pm}_{n/2}(p)$ are those resulting from the
combinations of the coexact and `exact' \zfs\ on ${\can}$ discussed above,
$$
\ze^{\can\pm}_{p}(s)=\ze^\can_p(s)\pm\ze^\can_{p-1}(s)
\eql{totzet3}$$

\section{\bf 4. The ball and sphere \zfs.}

We now consider forms on that cone whose base is
a unit $d$-sphere, $d=D-1$, \ie on the $D$-ball. This will be our main 
application of the general formalism exposed in 
the previous sections. As we have seen, it is sufficient to 
look at coexact forms (\ie transverse, antisymmetric fields).

The form \zfs\ (\peq{modcoexzet}) on the $d$-sphere are needed in order to 
find the corresponding
heat-kernel coefficients for substitution into the fundamental equation, 
(\peq{comancoeff}). The spectral properties have been known 
for some time [\pref{GandM,BandM,IandK,IandT,CandT,RandT}] and 
the coexact $p$-form eigenvalues of the de Rham Laplacian are readily 
established to be 
$$\mu(p,l)= \big(l+(d-1)/2\big)^2-\big((d-1)/2-p\big)^2,\quad l=1,2,\ldots.
\eql{eigvals}$$ 

As anticipated, we again witness the important simplification of the Bessel
function order, (\peq{reln}), to the $p$-independent form
$$
\nu(p,l)= l+(d-1)/2\,>(d-1)/2
\eql{nu}$$
exactly as in the scalar case. The first consequence is that the absolute
\zf\ (\peq{coexze}) simplifies to
$$
\ze^\ball_{a,p}(s)=\!\!\sum\!\int_\ga{dk\over2\pi i}k^{-2s}{\pa\over\pa k}\!
\bigg(\tilde d(p)\big(r^{u_a(p)}\ln J_\nu(kr)\big)'\bigg|_{r=1}
+d(p-1)\ln J_\nu(k)\bigg) .
\eql{ballzet}$$

For the time being let us work at a fixed $p$. The sphere coexact $p$-form 
degeneracy is 
$$
d(p,l)={(2l+d-1)(l+d-1)!\over p!(d-p-1)!(l-1)!(l+p)(l+d-p-1)}.
\eql{degen}$$
We note the symmetry, $d(p,l)=d(d-1-p,l)$ and that $d(d,l)=0$.
In addition, there is a zero mode for $p=0$ and one for $p=d$.

Rewrite (\peq{degen}) as
$$
d(p,l)={(l+d-1)!\over p!(d-p-1)!(l-1)!}\bigg({1\over l+p}+{1\over l+d-p-1}
\bigg),
\eql{degen2}$$
and consider, firstly, the sum 
$$
\sum_{l=1}^\infty{(l+d-1)!\over(l-1)!}{r^{l+(d-1)/2}\over l+p}
\eql{sum2}$$
with $r=\exp(-\tau)<1$. The idea is that this gives the 
square-root heat-kernel, and the sphere \zf, (\peq{modcoexzet}), follows by 
Mellin transform on $\tau$ as in our earlier works, [\pref{Dow1,ChandD}].

The generating function for a given form order can be rewritten using
the identity
$$
\sum_{l=1}^\infty{(l+d-1)!\over(l-1)!}{r^l\over l+p}=
(d-p-1)!\sum_{m=p+1}^d{(m-1)!\over(m-p-1)!}{r\over(1-r)^m}
\eql{iden}$$
which follows easily from recursion.\mgn{Checked SUMMS.MTH}

There is still an overall factor of $r^{(d-1)/2}$ in (\peq{sum2}) and 
performing the Mellin transform  produces a series of
Barnes \zfs, giving, after the addition of the $p\to d-p-1$ term, the 
{\it modified} coexact \zf, (\peq{modcoexzet}), on the sphere as
$$\eqalign{
\ze_p^{S^d}(s)=&
\sum_{m=p+1}^d\comb{m-1}p\ze_{\cal B}
\big(2s,(d+1)/2\mid{\bf1}_m\big)\cr
&+\sum_{m=d-p}^d\comb{m-1}{d-p-1}\ze_{\cal B}
\big(2s,(d+1)/2\mid{\bf1}_m\big)\cr}
\eql{spcoze}$$
for $0\le p<d$. Obviously $\ze_d^{S^d}(s)=0$.

(\peq{spcoze}) is formally much simpler 
than expanding the degeneracy to give a series of Hurwitz \zfs\ (a series 
noted in passing  by Copeland and Toms, [\pref{CandT}] and frequently 
used). This will come later.

When $p=0$, (\peq{spcoze}), with the zero mode included according
to (\peq{zetzeroin}), gives the known scalar expression. 
The first sum reduces to a term 
that is cancelled by the zero mode and to a single Barnes \zf, the result
being, [\pref{ChandD}], [\pref{BKD}] eqn. (5.7),
$$
\tilde\ze_0^{S^d}=\zb 2s; (d+1)/2\mid {\bf 1}_d) 
+\zb 2s; (d-1)/2\mid {\bf1}_d ).
\eql{0zeta}$$

This is elegantly seen from a rearrangement of (\peq{spcoze}), detailed
in Appendix A, that yields $\ze_p^{S^d}(s)$ as a finite series of scalar
\zfs\ by means of a recursion relating a $p$-form in $d$ dimensions to
a $(p-1)$-form in $d-2$. The series is ($0\le p<(d-1)/2$)
$$\eqalign{
\tilde\zeta_p^{S^d} (s) =\sum_{j=0}^p &(-1)^j \left( d-1-2j \atop p-j 
\right) 
\tilde\ze_0^{S^{d-2j}}(s)\cr
&+\de_{p,0}\bigg({d-1\over2}\bigg)^{-2s}
-(-1)^p\bigg({d-2p-1\over2}\bigg)^{-2s}\cr}
\eql{finser}
$$
and this is one of our basic equations. Duality, (\peq{zedual}), can be 
used to extend the range of $p$.

When $d$ is odd we have from (\peq{0zeta}) the special values for
$q\in\oZ$,
$$
\tilde\ze_0^{S^d}(-q)=0,\quad q\ge0,
\eql{qval0}$$
which corresponds to the fact that for scalars the relevant
operator is the (improved) geometrical Laplacian. 

From (\peq{qval0}) follow the special values of the modified {\it coexact} 
function \mgn{GENBERN.MTH} ($0<p<d$)
$$
\ze_p^{S^d}(-q)=\tilde\ze_p^{S^d}(-q)=
(-1)^{p+1}\big((d-1)/2-p\big)^{2q},
\eql{spval}$$
In particular
$$
\ze_p^{S^d}(0)=(-1)^{p+1},
\eql{0val}$$ 
and
$$
\ze_{(d-1)/2}^{S^d}(-q)=0,\quad q>0.
\eql{qval}$$

Hodge duality, as expressed in (\peq{zedual}), can be seen explicitly from 
(\peq{spcoze}). Inclusion of the zero modes obviously violates this duality.

It is also worth pointing out that the + combination of coexact and 
exact \zfs\ in (\peq{totzet3}) on the sphere, in contrast to 
(\peq{totzet2}) 
on the ball, does not correspond to the \zf\ for any Laplacian.

Finally, we note that $\ze_p^{S^d}(s)$ does not have a pole at 
$s=-1/2$, which is important during the construction of the heat-kernel 
coefficients and the functional determinants on $\man$. It means that the
$\log$ term is absent in the expansion (\peq{expans}), \cf [\pref{BKD}].
No further discussion of this term is given here. Some relevant comments
can be found in the recent works by Bytsenko {\it et al} 
[\pref{BCZ1,BCZ2}].

\section{\bf 5. Residues and sphere heat-kernel coefficients.}
To apply (\peq{comancoeff}) to the ball it is necessary to know the 
coefficients on the sphere. We use (\peq{coefff}) on $\can$ with the 
appropriate \zf, in this case either $\ze^\can$ or $\tilde\ze^\can$.

From (\peq{finser}), the residue of the modified {\it coexact} sphere 
\zf\ at $s=k/2$, $k\in\oZ$, is
$$
\Res\ze_p^{S^d}(k/2)={1\over(k-2)!}\sum_{j=0}^p(-1)^j
{2^{k-d+2j}D^{(d-2j-1)}_{d-2j-k}
\over (d-2j-1)(d-2j-k)!}\comb{d-2j-1}{p-j}
\eql{resi2}$$
where we have been able to use the results of our earlier work 
[\pref{BKD}] for the residues of 0-form \zfs. The $D^{(n)}_\nu$ are easily 
evaluated generalised Bernoulli polynomials.

The corresponding heat-kernel coefficients are, for ${\rm min}\,([n/2],p,d-p)=
[n/2]$ with $d-k=n$,
$$
{(4\pi)^{d/2}\over|S^d|}A^{S^d}_{n/2}(p)=
(k-1){\Ga\big((d+1)/2\big)\over\Ga\big((k+1)/2\big)}
\sum_{j=0}^{[n/2]}{(-1)^j2^{2j}
D^{(d-2j-1)}_{n-2j}\over(d-2j-1)(n-2j)!}\comb{d-2j-1}{p-j}.
\eql{co5}$$
$A^{S^d}_{n/2}(p)$ vanishes when $n$ is odd and less than $d$ because
$D^{(n)}_\nu$ is zero for odd $\nu$.

A useful check is provided by the modified coexact coefficients 
$A^{S^d}_{q+d/2}=(-1)^q\ze_p^{S^d}(-q)/q!$, $q\in \oZ^+$, for odd $d$. 
According to (\peq{spval}) these half-odd coefficients are non-zero, in
agreement with the results of [\pref{ELV1}] and [\pref{CandH2}] for the de 
Rham coexact \zf. 
For this case, the only half-odd coefficient is the $q=0$ one, 
corresponding 
to $\ze(0)$, and we know that the modified and de Rham coexact heat-kernels 
are related by a factor,
$$
K_p^{\rm mod}(\tau)=e^{-\big((d-1)/2-p\big)^2\tau}\,K_p^{\rm de Rham}(\tau),
\eql{hkreln}$$
the expansion of which reproduces exactly (\peq{spval}). There are no 
half-odd coefficients for $n<d$, nor for $p=0$ when the zero mode is
included.

Similar statements hold for the `total' coefficient, $A^{S^d+}_{n/2}$. 
Because of the construction, (\peq{totzet3}), of $\ze_p^{S^d+}$, one 
should not, of course, expect a standard geometric form, except when $p=0$.

As a curiosity, constructing the `total' sphere coefficient 
(\cf (\peq{totzet3})) by just adding equation (\peq{co5}) for $p$ and 
$p-1$, we have
$$
{(4\pi)^{d/2}\over|S^d|}A_{n/2}^{S^d+} = 
(k-1){\Ga\big((d+1)/2\big)\over\Ga\big((k+1)/2\big)}
\sum_{j=0}^{[n/2]} (-1)^j 
\frac{2^{2j-n} D_{n-2j}^{(d-2j-1)}}
{(d-2j-1) (n-2j)!}\left( d-2j\atop p-j \right),
\eql{sphere1}
$$
similar to the general form on the ball developed in the
next section and also in agreement with the results of G\"unther and
Schimming [\pref{GandS}] and Gilkey, \eg [\pref{Gilkey1}].

Equation (\peq{hkreln}) can be used, in familiar fashion, to give general 
formulae for the standard,\ie de Rham, coexact heat-kernel coefficients on 
the sphere, denoted $B_{n/2}$, somewhat simpler than those derived in 
[\pref{CandH2}] and [\pref{ELV1}]. Thus
$$\eqalign{
B^{S^d}_M(p)&=\sum_{N=0}^M {w(p)^{M-N}\over(M-N)!}\,A_N^{S^d}(p)\cr
B^{S^d}_{M+1/2}(p)&=\sum_{N=0}^M {w(p)^{M-N}\over(M-N)!}\,
A_{N+1/2}^{S^d}(p)\cr}
\eql{deRcoeffs}$$
where $w(p)=\big((d-1)/2-p\big)^2$ and $A_{n/2}^{S^d}(p)$ is given by 
(\peq{co5}). For odd $d$, the only non-zero half-odd coefficient is
$B^{S^d}_{d/2}(p)$, as has already been remarked.

Elizalde {\it et al} [\pref{ELV1}] give selected numerical values for the 
total $B$ coefficients obtained from a general formula that involves a 
series of Hurwitz \zfs\ and is slightly less convenient than the above. 
We have found agreement with their values. 

The connection between the two expressions is simply that the result of 
[\pref{ELV1}] follows from ours if the generalised Bernoulli polynomials
are expanded in ordinary ones. From a numerical point of
view this is unnecessary because there are efficient algorithms that 
give the generalised functions directly. 

\section{\bf 6. General form of coefficients on the ball.}
We now come to the central part of this paper, the explicit construction
of the heat-kernel coefficients on the ball.

Firstly we can think of checking against the limited results of 
[\pref{BBG}] for the coefficients for the form--valued 
Laplacian on manifolds with boundary.

From Theorem 1.2 of [\pref{BBG}] applied to the $D$-ball, 
the {\it total} coefficient expressions for absolute boundary conditions 
are,
(the $+$ superscript is dropped in this section)
$$\eqalign{
{(4\pi)^{d/2}\over|\rS^d|}A^{(D)}_0(p)&={1\over2\sqrt\pi}{1\over(d+1)}
\comb Dp\cr
{(4\pi)^{d/2}\over|\rS^d|}A^{(D)}_{1/2}(p)&=
{1\over4}\bigg[\comb Dp-2\comb {D-1}{p-1}\bigg]\cr
{(4\pi)^{d/2}\over|\rS^d|}A^{(D)}_{1}(p)&=
{1\over\sqrt\pi}{d\over6}\bigg[\comb Dp-6\comb {D-1}{p-1}
+6\comb{D-2}{p-2}\bigg]\cr
{(4\pi)^{d/2}\over|\rS^d|}A^{(D)}_{3/2}(p)&=
{d\over384}\bigg[(13d+2)\comb Dp-4(29d-26)\comb {D-1}{p-1}+\cr
&\hspace{******}288(d-1)\comb{D-2}{p-2}-192(d-1)\comb{D-3}
{p-3}\bigg].\cr}
\eql{BBG12}$$

Assuming, according to our general philosophy, that
$d$ is sufficiently large,
it has been checked that the first three coefficients follow from
our formulation using (\peq{comancoeff}), (\peq{co5}) and various 
binomial identities including the recursion
$$
\comb ab=\comb {a-1}b+\comb{a-1}{b-1}.
\eql{brecurs}$$
The fourth coefficient will be confirmed later by a
more powerful technique, and the series continued.

Although these results were derived on the basis of a restriction  on
$p$, no such restriction exists for the expressions 
(\peq{BBG12}) in [\pref{BBG}]. The conclusion is that, once the 
coefficients are 
displayed as explicit functions of $d$ and $p$, they can be continued 
outside 
the restrictions, the reason being that their particular structure is 
guaranteed by the geometric formulation. 

These particular checks show that our general formulae are correct 
but it is not feasible to pursue the further evaluation of the coefficients
in this way by algebraic rearrangement of the produced expressions, since 
it involves awkwardly repeated application of the binomial recursion and of 
other identities. 
A systematic approach to the evaluation of any coefficient is better provided 
by fitting unknowns in a general form, and this process will now be set in 
train.

From basic theory (\eg [\pref{Gilkey1}]) the geometric expression on a 
flat, 
bounded $D$-manifold, $\man$, is, up to terms involving derivatives of the 
extrinsic curvature $\bka$,
$$
c(n)(4\pi)^{d/2}A^D_{n/2}(p)=\int_{\pa\man}b_{\bf n}(D,p)\sum_{\bf n}
\big(\tr(\bka^{n_1})\tr(\bka^{n_2})\ldots\big)
\eql{flgen}$$
with
$$\eqalign{
c(n)&=2\sqrt\pi,\quad n\,\, {\rm even}\cr
&=1,\quad n\,\,{\rm odd}\cr
&=2(d+1)\sqrt\pi,\quad n=0\cr}
$$
and where ${\bf n}=(n_1,n_2,\ldots)$ is a partition of $n-1$. For 
convenience
the $n=0$ term has been included although it is really a volume 
contribution.
We seek the $b_{\bf n}(D,p)$.

For the $D$-ball, (\peq{flgen}) reduces to
$$
c(n)(4\pi)^{d/2}A^D_{n/2}(p)=|\rS^d|\,\sum_{\bf n}d^{|{\bf n}|}
b_{\bf n}(D,p)=|\rS^d|\,\sum_{k=1}^{n-1}d^k\,b_k^{(n)}(D,p)
\eql{ballg}$$
where $|{\bf n}|$ is the number of components in the partition and 
$b_k^{(n)}$,
is the sum of those $b_{\bf n}$ for which $|{\bf n}|=k$, 
$$
b_k^{(n)}(D,p)=\sum_{\bf n}b_{\bf n}(D,p)\bigg|_{|{\bf n}|=k}.
$$
These are the only combinations that can be determined from working on 
the ball.

The numerical multipliers $b_{\bf n}(D,p)$ satisfy the binomial recursion
$$
b_{\bf n}(D,p)=b_{\bf n}(D-1,p)+b_{\bf n}(D-1,p-1),
\eql{bbrec} $$
proved by crossing $\man$ with a unit circle, [\pref{BBG}]. 
This relation is what has become of the more familiar statement of 
dimension-independence for scalars.

Thus $b_{\bf n}(D,p)$ can be expanded as a linear combination of binomial
coefficients $\comb{D+a}{p+b}$ for varying $a$ and $b$. Since 
$b_{\bf n}(D,p)$ vanishes for $p$ outside the range $0$ to $D$, 
$a$ must equal $b$ and be nonpositive. Consequently the expansion reads
$$
b_{\bf n}(D,p)=\sum_{m=0}^nM_{{\bf n},m}\comb{D-m}{p-m}
\eql{bexp}$$
which is the boundary version of the G\"unther and Schimming form 
[\pref{GandS}].
The limits have been fixed by noting that they must be independent of 
both $D$ and $p$ and, therefore, can be set by considering the particular 
value $D=n$, for which there are $n+1$ independent constants implying $n+1$
terms in the sum. The trivial first coefficient, $b_0=\comb D p$, can 
also be used to pin down the limits to those shown. 
The conclusion is that, on the ball, the general form can be written,
$$
{c(n)(4\pi)^{d/2}\over|\rS^d|}A^{(D)}_{n/2}(p)=
\sum_{m=0}^nP_m^{(n)}(d)\comb{D-m}{p-m}
\eql{genformcoeff}$$where
$P_m^{(n)}(d)$ is a polynomial of degree $n-1$ in $D$. 
For $n>1$
$$
P_m^{(n)}(d)=\sum_{\bf n}M_{{\bf n},m}d^{|{\bf n}|}
=\sum_{k=1}^{n-1} M^{(n)}_{k,m}\,d^k
\eql{ppoly}$$where the $M^{(n)}_{k,m}$ are constants. 

Our method is simply that, for fixed $n$, knowing the left-hand side 
for $0\le p\le n$, (\peq{genformcoeff}) can be inverted for the 
$P_m^{(n)}(d)$, keeping $d$ unspecified. The method employed in 
[\pref{BBG}] uses the particular values $b_{\bf n}(n,p)$ to invert
(\peq{bexp}). Our technique is more flexible in practice, although the 
analysis in [\pref{BBG}] does bring out clearly the amusing fact that 
the general coefficient 
can be generated from just the values $A^{(n)}_{n/2}(p)$, $0\le p\le n$.

In detail, we firstly note that, for $p=0$ to $n$, the matrix of binomial 
coefficients on the right-hand side of (\peq{genformcoeff}) is triangular
so that the inversion follows recursively by forward substitution,
$$
P_m^{(n)}(d)={c(n)(4\pi)^{d/2}\over|\rS^d|}A_{n/2}^{(D)}(m)-
\sum_{\mu=0}^{m-1}\comb {D-\mu}{m-\mu}P^{(n)}_\mu(d),\quad m=0,\ldots, n.
\eql{fsub}$$

The driving coefficients $A_{n/2}^{(D)}(m)$ ($0\le m\le n$) will be 
determined from (\peq{comancoeff}) as polynomials in $d$ because, 
for given numerical values of $p$ and $n$, the sphere coefficients, 
(\peq{co5}), are obviously such polynomials.

It should be stated that the \zf\ on the ball is a standard one, and that 
the heat-kernel coefficients are geometric (leading to the general form 
(\peq{genformcoeff})). In this case it is quite permissible to evaluate the
polynomials in whatever parameter region is convenient and then continue, 
despite the fact that they involve coefficients on the sphere for which 
this might not be possible.

Evaluation is a routine machine matter, the first outcome
of which is $A^{(D)}_{3/2}(p)$ in (\peq{BBG12}). Some results are
given below in the form of matrices of the constants $M^{(n)}_{k,m}$ in 
(\peq{ppoly}), \mgn{MMATR.MTH}
$${\bf M}_a^{(3)}=
\left(\matrix{{2\over192}&{13\over48}&-{3\over4}&-{1\over2}\cr
{13\over384}&-{29\over96}&{3\over4}&1\over2\cr}\right)
\eql{al3}
$$
\vskip15truept
$${\bf M}_a^{(4)}=
\left(\matrix{{4\over{135}}&-{{164}\over{315}}&{{16}\over5}&-{{16}\over 3}&
{8\over3}\cr
{1\over{45}}&{{92}\over {105}}&-{{74}\over {15}}& {\scriptstyle8}& 
{\scriptstyle-4}\cr
{1\over{27}}&-{{136}\over{315}}&{{26}\over{15}}&-{8\over3}&{4\over 3}\cr}
\right)
\eql{al4}$$
\vskip15truept
$${\bf M}_a^{(5)}=
\left(\matrix{
{{77}\over {15360}}&
{{77}\over {960}}&
 -{{191}\over {192}}&
{{19}\over 6}&
 -{{15}\over 4}&
{3\over 2}\cr
{{235}\over {36864}}&
-{{263}\over {1440}}&
{{1475}\over {768}}&
 -{{47}\over 8}&
{{55}\over 8}&
-{{11}\over 4}\cr
{{139}\over {61440}}&
{{1987}\over {15360}}&
-{{1769}\over {1536}}&
{{157}\over {48}}&
-{{15}\over 4}&
{3\over 2}\cr
{{2041}\over {737280}}&
-{{3787}\over {92160}}&
{{347}\over {1536}}&
-{9\over {16}}&
 {5\over 8}&
 -{1\over 4}\cr}\right).
\eql{al5}$$
\vskip15truept

A subscript has been added to indicate that these values are for absolute
boundary conditions.
Although any relative coefficient can be found from the absolute ones
using duality, in order to obtain the general form
the entire analysis should be repeated. A tactically better way combines 
these procedures and consists
of using the recursive nature of the relative coefficients to write
down a general form, precisely as in (\peq{genformcoeff}), together with the 
corresponding solution, (\peq{fsub}), but now where the values of the
coefficients for $p=0$ to $n$ are determined by duality in terms
of already evaluated absolute quantities.  In any case, one rapidly finds 
that the first coefficient is the same, that the second is reversed in sign
and that the remaining ones (up to $n=5$) are contained in the matrices,
$$
{\bf M}_r^{(3)}=\left(\matrix{ 
{5\over {192}}&
-{13\over {48}}&
{3\over 4}&
-{1\over 2}\cr
-{7\over {384}}&
{{29}\over {96}}&
 -{3\over 4}&
{1\over 2}\cr}
\right)
$$

$$
{\bf M}_r^{(4)}=\left(\matrix{
{8\over {189}}&
-{{172}\over {315}}&
{{16}\over 5}&
-{{16}\over 3}&
{8\over 3}\cr
-{{11}\over {315}}&
{{104}\over {105}}&
-{{74}\over {15}}&
{\scriptstyle8}&
{-\scriptstyle4}\cr
{1\over {189}}&
-{{116}\over {315}}&
{{26}\over {15}}&
-{8\over 3}&
{4\over 3}\cr}
\right)
$$
\vskip10truept
$$
{\bf M}^{(5)}_r=\left(\matrix{
{{109}\over {15360}}&
-{{29}\over {320}}&
{{193}\over {192}}&
-{{19}\over 6}&
{{15}\over 4}&
-{3\over 2}\cr
-{{1049}\over {184320}}&
{{1247}\over {5760}}&
-{{1501}\over {768}}&
{{47}\over 8}&
-{{55}\over 8}&
{{11}\over 4}\cr
{{47}\over {61440}}&
-{{709}\over {5120}}&
{{1783}\over {1536}}&
-{{157}\over {48}}&
{{15}\over 4}&
-{3\over 2}\cr
{{13}\over {147456}}&
{{2467}\over {92160}}&
-{{325}\over {1536}}&
{9\over {16}}&
-{5\over 8}&
{1\over 4}\cr}
\right).
\eql{rem5}$$
\vskip10truept
\noin It is a matter of a few minutes by machine algebra to calculate 
larger matrices.

Knowledge of these constants enables restrictions to be placed on the 
multipliers, $b_{\bf n}(D,p)$, in the general geometric form of the 
coefficients, (\peq{flgen}).

Denoting the set of $b_k^{(n)}(D,p)$, $(1\le k\le n-1)$,
by the vector ${\bf b}^{(n)}$, the vector of binomial coefficients
$\comb {D-m}{p-m}$, $(0\le m\le n)$, by ${\bf c}^{(n)}$ and the set of
powers $d^k$ ($1\le k\le n-1$) by $\bf d^{(n)}$, equation 
(\peq{genformcoeff}) takes the matrix form
$$
{c(n)(4\pi)^{d/2}\over|\rS^d|}A^{(D)}_{n/2}(p)=\widetilde{{\bf d}^{(n)}}
{\bf M}^{(n)}{\bf c}^{(n)}
\eql{coefmat}$$
and the restrictions are
$$
{\bf b}^{(n)}={\bf M}^{(n)}{\bf c}^{(n)}.
\eql{rest}$$

For example, for $n=5$ there are 4 partitions so
$$
b_1^{(5)}=b_{4},\quad
b_2^{(5)}=b_{1,3}+b_{2^2},\quad
b_3^{(5)}=b_{1^2,2},\quad
b_4^{(5)}=b_{1^4}
$$
and only these particular combinations are known using (\peq{rest}) and
(\peq{al5}) or (\peq{rem5}).
It is clear that one can always unambiguously determine $b^{(n)}_{n-1}$ and 
$b^{(n)}_{1^{n-1}}$.

We finally note that the $n=4$ values at $D=4$ agree with those in Table 2 
of Moss and Poletti [\pref{MandP}].\mgn{MMATR.MTH}
\section{\it Generating functions.}
A generating function approach to the form order, $p$, could be introduced 
at this coefficient stage by defining 
$$
\overline A_{n/2}^{(D)}(\si)=
\sum_{p=0}^D(-\si)^p A^{(D)}_{n/2}(p)
={|\rS^d|\over c(n)(4\pi)^{d/2}}\sum_{m=0}^n(-\si)^m(1-\si)^{D-m}P_m^{(n)}
(d).
\eql{genfcoeff}$$

In particular the value $\si=-1$ is required (\cf [\pref{BBG}] Thm. 1.3),
$$
\overline A_{n/2}^{(D)}(-1)
={4\sqrt\pi\over c(n)\Ga\big((d-1)/2\big)}\sum_{m=0}^n2^{-m}P_m^{(n)}(d)
\eql{sumcoef}$$
which is a convenient way of organising and computing this quantity, at 
this point.

There is topological interest in the case $\si=1$ and we see
from (\peq{genfcoeff}) that
$$
\overline A_{a,n/2}^{(D)}(1)=0,\quad n<D;\quad
\overline A_{a,D/2}^{(D)}(1)={(-1)^D2^{1-d}\sqrt\pi
\over c(D)\Ga\big((d-1)/2\big)}P^{(D)}_D(d)=\chi(\man)=1
\eql{euler}$$
which provides a check of the Gauss-Bonnet theorem. If relative
coefficients are used we find $\chi(\man,\pa\man)=-1$, which is correct.

\section{\bf 7. Value at zero.}
Special interest, both mathematical and physical, attaches itself to the 
particular value 
$\ze^{\man+}_p(0)$ which merits individual treatment, \eg
[\pref{MandP, Esposito, Barv}]. Further, 
as already said, these values play an 
important role in the algebraic computations of Bla\v zi\'k {\it et al} 
[\pref{BBG}].

Generally, for absolute conditions, the coexact expression is
$$\eqalign{
\ze^{\man}_p&(0)=-{1\over2}\big(\tilde\ze^{\can}_{p}(-1/2)
+\ze^{\can}_{p-1}(-1/2)\big)+{1\over4}\big(\tilde\ze^{\can}_{p}(0)-
\ze^{\can}_{p-1}(0)\big)\cr
&-\sum_{k=1}^d{1\over k}\bigg(\ze_R(-k)\big(\Res\ze^{\can}_{p}(k/2)
+\Res\ze^{\can}_{p-1}(k/2)\big)
-\big(-u_a(p)\big)^k\Res\ze^{\can}_{p}(k/2)\bigg).\cr}
\eql{0val2}$$

For $\can=\rS^d$ the residues are contained in (\peq{resi2}) and the 
values of the \zf\ found from (\peq{spcoze}) or (\peq{finser})
(there are no zero modes on the sphere {\it in the sense that} $\nu>0$).

An alternative evaluation consists of using the standard formula
$$
\ze_p^{\man+}(0)=A^{(D)}_{D/2}(p)-\be^\man_p
\eql{standeq}$$
together with an explicitly computed general coefficient 
(\peq{genformcoeff}). 
The absolute Betti numbers of the ball are $\be_0=1$, the rest vanishing.

The direct evaluation via (\peq{0val2}) is the most efficient, numerically. 
Machine computation shows agreement between these two approaches and also
with the values presented in [\pref{ELV2}], except that the $p=0$ numbers
for $d=6$ and $7$ differ.

It is a remarkable fact that in using (\peq{comancoeff}), leading to
(\peq{genformcoeff}) and thence to (\peq{standeq}), one does not have to worry
about the `zero modes' in say (\peq{zetzeroin}), whilst such contributions 
are vital when evaluating directly from (\peq{0val2}).

Finally, using the inverse (\peq{inver}) and noting that $\ze^\man_D(0)$ 
vanishes, we reconfirm the topological formula
$
\chi=\sum_{p=0}^D(-1)^p\,\be^\man_p=\sum_{p=0}^D(-1)^p\, A^D_{D/2}(p).
$
\section{\bf 8. Form functional determinants.}

In this section we consider functional determinants for $p$-forms. As 
we have seen in the calculation of the heat-kernel coefficients, it will be
enough to consider the determinant associated with the coexact \zf\
(\peq{coexze}), this being simply a combination of Robin and
Dirichlet contributions. It is immediately appreciated that the results 
derived in [\pref{BKD}], eq.~(9.8) for Dirichlet boundary conditions
and  eq.~(11.2) for Robin conditions, remain valid once the base \zf\
there is replaced with the $p$-form base \zf, 
eq.~(\peq{modcoexzet}), with, in addition,
$$
\zeta^{{\cal N}+1}_p (s) = \sum_{n=1}^{\infty}\sum 
\frac{d(p)}{\big(\nu (p) +n\big)^{s}},\,\,\,\,{\rm and}\,\,\,\,\,
\zeta^{{\cal N}}_p (s,u) = \sum
\frac{d(p)}{\big(\nu (p) +u\big)^{s}},
\eql{zets}$$
and the related quantities, ${\tilde \zeta} _p ^{{\cal N} +1} (s)$, 
${\tilde \zeta} _p ^{{\cal N}} (s,u)$, if the zero modes are included as 
in (\peq{zetzeroin}).

The coexact determinant for $p\geq 1$
combines Robin with Dirichlet and is determined, for absolute conditions, by
the derivatives,
$$\eqalign{
{\zeta'}^{\man}_p(0)=&{\tilde \zeta}^{\prime\can+1}_p(0)
+{\zeta '}^{\can +1}_{p-1}(0)+
{\tilde \zeta}^{\prime\can} _p\big(0,u_a (p)\big)
+\ln 2\bigg({\tilde \zeta} ^{\can}_p (-1/2)
+\zeta^{\can}_{p-1}(-1/2)\cr
&+2\sum_{i=1\atop i\,\,odd}^d\Res\zeta_p^{\can} 
(i/2)\, M_i\big(1,u_a(p)\big)+2\sum_{i=1}^d\Res\zeta_{p-1}^{\can} (i/2)\, 
D_i(1)\bigg)\cr
&+2\sum_{i=1 \atop i\,\, odd}^d \Res\zeta_p^{\can} (i/2) 
\bigg( M_i\big(1,u_a(p)\big)\sum_{k=1}^{i-1} 1/k \cr
&\hspace{*********}+\int_0^1 dt\,\,
\frac{M_i\big(t,u_a (p)\big)-tM_i\big(1,u_a(p)\big)}{t(1-t^2)}\bigg)\cr
&+2\sum_{i=1 \atop i\,\, even}^d \Res\zeta_p^{\can} (i/2) 
\bigg( M_i\big(1,u_a(p)\big) \sum_{k=1}^{i-1} 1/k \cr
&\hspace{*********}+\int_0^1 dt\,\,
\frac{M_i\big(t,u_a (p)\big)-t^2 M_i\big(1,u_a(p)\big)}{t(1-t^2)}\bigg)\cr
&+2\sum_{i=1}^d \Res\zeta_{p-1}^{\can} (i/2) 
\bigg(D_i (1)\sum_{k=1}^{i-1} 1/k +\int_0^1 dt\,\,
\frac{D_i (t)-tD_i (1)}{t (1-t^2)}\bigg).}
\eql{det}$$
The $M_i$ and $D_i$ are known polynomials associated with the asymptotic
expansion of Bessel functions and, to avoid repetition, we refer to 
[\pref{BKD}], and to references therein, for more details. Some information 
can be found incidentally in Appendix D. 

For $p=0$ a small extra consideration is necessary. As we have mentioned,
absolute 0-forms are Neumann scalars
and the Robin parameter is then $u_a (0) = -(d-1)/2$. Looking 
at $\zeta_p ^{\can} (s,u)$ in eq.~(\peq{zets}) it is seen that, for 
$u=-\nu (p)$, one encounters a branch cut occasioned by (incorrectly) 
including the true zero mode for Neumann conditions. The technical 
reason is that the asymptotic expansion for these
specific Robin parameters is slightly different from the others.

The easiest way of taking this into account is to subtract the contribution
in (\peq{zets}) for $u\neq -\nu (p)$, then take the limit as $u\to 
-\nu (p)$ and finally to add the correct contribution for $u=-\nu (p)$. 

For Neumann conditions the end result is
$$\eqalign{
{\zeta'}^{\man}_0 (0) =&{\tilde \zeta}^{\prime\can+1}_0 (0) 
+\ln 2\bigg(\tilde \zeta^{\can}_0 (-1/2)
+2\sum_{i=1\atop i\,\,odd}^d\Res\zeta_0^{\can}(i/2)\, M_i\big(1,u_a(0)\big)
\bigg)  \cr
\noa
&+2\sum_{i=1 \atop i\,\, odd}^d \Res\zeta_0^{\can} (i/2)
\bigg( M_i\big(1,u_a(0)\big)\sum_{k=1}^{i-1} 1/k \cr
\noa
&\hspace{*********}+\int_0^1 dt\,\,
\frac{M_i\big(t,u_a (0)\big)-tM_i\big(1,u_a(0)\big)}{t(1-t^2)}\bigg)\cr
\noa
&+2\sum_{i=1 \atop i\,\, even}^d \Res\zeta_0^{\can} (i/2)
\bigg( M_i\big(1,u_a(0)\big) \sum_{k=1}^{i-1} 1/k \cr
\noa
&\hspace{*********}+\int_0^1 dt\,\,
\frac{M_i\big(t,u_a (0)\big)-t^2 M_i\big(1,u_a(0)\big)}{t(1-t^2)}\bigg)\cr
\noa
&+\lim_{u\to -(d-1)/2 } \left(
{\tilde \zeta _0} ^{\prime\can} (0,u) +\ln \left( 
(d-1)/2+ u \right)\right)
+\ln \big(d+1\big). \cr}
\eql{detneu} 
$$
The relative, Dirichlet expression is given in [\pref{BKD}].

Equations (\peq{det}) and (\peq{detneu}) are expressions on the generalised
cone and we apply them now to the ball, 
for which all quantities have already been treated, apart from 
$\zeta^{{\cal N}+1}_p (s)$ 
($\tilde \zeta^{{\cal N}+1}_p (s)$)
and $\zeta^{{\cal N}}_p (s,u)$
($\tilde \zeta^{{\cal N}}_p (s,u)$).
For these remaining \zfs\ one immediately finds, along the lines of 
[\pref{BKD}],
$$\eqalign{
\tilde \zeta^{{\cal N}+1}_p (s)=&    
\sum_{m=p+1}^d\comb{m-1}p\ze_{\cal B}
\big(s,(d+3)/2\mid{\bf1}_{m+1}\big) 
+\ze_R\big(s; (d+3)/2)\big)\de_{pd}
\cr
&+\sum_{m=d-p}^d\comb{m-1}{d-p-1}\ze_{\cal B}
\big(s,(d+3)/2\mid{\bf1}_{m+1}\big)
+ \ze_R\big(s; (d+1)/2)\big)\de_{p0}\cr}
\eql{zetu1}$$
and
$$\eqalign{
\tilde \zeta^{{\cal N}}_p (s,u_a) = &\sum_{m=p+1}^d\comb{m-1}p\ze_{\cal B}
\big(s,p+1\mid{\bf1}_{m}\big)\cr
&+\sum_{m=d-p}^d\comb{m-1}{d-p-1}\ze_{\cal B}
\big(s,p+1\mid{\bf1}_{m}\big).\cr
\noa
&\hspace{****}+\delta_{p0}\big((d-1)/2+u\big)^{-s} 
+\delta_{pd} (d+1)^{-s}.\cr}
\eql{zetu}$$
The contribution of the zero modes is visible and 
$ \zeta^{{\cal N}+1}_p (s)$, $ \zeta^{{\cal N}}_p (s,u_a) $ need not
be stated. It is clearly seen that the limit $u\to-(d-1)/2$ in 
(\peq{detneu}) is well defined because the logarithm is cancelled by the
last term in (\peq{zetu}). 

Using (\peq{zetu1}) and (\peq{zetu}) in 
(\peq{det}) and (\peq{detneu}), the determinants emerge as derivatives 
of the Barnes \zf\ at $s=0$. These may be expanded using Stirling
numbers,  
$$\eqalign{
\zeta_{{\cal B}} \big(s,a\mid {\bf1}_m) &= \sum_{n=0}^{\infty} 
\comb{m+n-1}m (a+n)^{-s} \cr
\noa
&= \sum_{k=1}^{m}{(-1)^{k+m}\over
(k-1)!(m-k)!}B^{(m)}_{m-k}(a)\,\zeta_R (s+1-k,a) \cr}
\eql{deg}$$
to give derivatives of Hurwitz or Riemann \zfs, [\pref{Barnesb}] p.433. 
The derivatives presented in Appendix B were obtained in this way. 
It is easy to find the expression in any dimension $d$ and for
any value of $p$. 

The presentation of our results is
a little more explicit than the corresponding ones of 
[\pref{ELV2}] in that only derivatives of Riemann \zfs\ are 
involved, there being no integrals over $\Gamma$ or $\Psi$ functions.
In fact, the integrals in [\pref{ELV2}] can be done, 
bringing exact agreement between the two sets of expressions.
This agreement is elaborated, in its essentials, in Appendix C.

It should be noted that, as a consequence of an earlier error in 
[\pref{BGKE}] for Neumann 0--forms, a $\log2$ term should be added to the 
results in [\pref{ELV2}] for $p=0$ and $d\ge2$.

\section{\bf 9. Spectral properties of spinors.} 
For the sake of completeness, the method of [\pref{BKD}] is now employed
to put the results of the earlier work on spinors, [\pref{DABK}], on a 
secure and systematic footing. For any undefined quantities and for further 
explanations, refer to this paper.
\vskip 10truept
\noin{\it Spectral boundary conditions.}
\vskip10truept
In [\pref{DABK}] it was shown that spectral boundary
conditions lead to 
$$
J_{n+ D/2 -1} (k) =0 ,\qquad n\in \oZ,\eql{1} 
$$
with the degeneracy
$$
d(n) =d_S \,\comb{n+D-2}n.\eql{2}
$$
Having in mind the treatment of the scalar field in [\pref{BKD}] we define 
the spinor base \zf\ appropriate for the ball,
$$
\zn s\right) = \sum_{n=0}^\infty d(n) \left(n+D/2 -1\right)^{-2s}.\eql{3}
$$
This choice guarantees that all formulas in terms
of $\zeta_{\can}$ in section 3 of [\pref{BKD}] remain valid. Also the 
polynomials $D_i (t)$ defined there are unchanged. In particular, eq.~(4.8) 
is still true, the only difference is in the base \zf\ used for 
the calculation of $A_{n/2}^{\can}$, the base heat-kernel coefficients.

Let us continue with the treatment of $\zn s\right)$ for the ball. It has 
the especially simple form
$$
\zn s\right) =d_S\, \zb 2s, (d-1)/2\mid{\bf 1} _d ).\eql{4}
$$
To obtain the heat-kernel coefficients using eq.~(4.8) of [\pref{BKD}], 
we need only the residues in (\peq{4}). 
Using the known values of the Barnes \zf\ in eq.~(4.8) of [\pref{BKD}] one 
regains all the results of [\pref{DABK}] for the heat-kernel coefficients 
including the conjecture for their general form. 

The determinants come out similarly.
Due to the identical structure of the present problem, eq.~(9.8) of 
[\pref{BKD}] remains valid once the base zeta
function there is replaced by (\peq{3}). The only `unknown' quantity is
$$
\ze_{\can +1} (s) =d_S\, \zb s,(d+1)/2 \mid{\bf 1} _{d+1}).\eql{7}
$$
Using (\peq{deg}) this can be expanded in Hurwitz-Riemann \zfs, and 
complete agreement with the results exhibited in [\pref{DABK}] is again
quickly and economically found.

In Appendix D we prove the curious identity for the spectral heat-kernel 
coefficients,
$$
A_{(D-2)/2}^{(S)}(D) =0, 
\eql{siden1}$$
observed numerically in I, the significance of which has not yet been 
discovered.

\section{\it Local boundary conditions.}
For local boundary conditions a few more technical details are necessary. 
The eigenvalue condition is, [\pref{DABK}],
$$
J_{n+D/2}^2 (k) -J_{n+D/2 -1} ^2 (k) =0,\quad n\in\oZ\eql{9}
$$
with the mode degeneracy 
$$
d(n) ={d_S\over2}\,\comb{n+D-2}n, \eql{10}
$$
which is half that of the spectral case. Defining the base 
\zf\ as in (\peq{3}) we have
$$
\zn s\right) =\frac 1 2 d_S\, \zb 2s,D/2 -1\mid{\bf 1}_d ).\eql{11}
$$
The polynomials describing the asymptotics are a bit different,
however, and read
$$
D_i (t) =\sum_{a=0}^{2i} x_{i,a} t^{i+a},  
$$
which leads to a slightly different form of the $A_i (s)$. These are 
found to be (one only has to refer to [\pref{KandC}] eq.~(2.18), and
replace the \zfs\ by the base \zf),
$$\eqalign{
A_{-1} (s) =& \frac 1 {2\sqrt{\pi}} \frac{\Gamma (s-1/2)}{\Gamma (s+1)}
\zn s-1/ 2 \right), \cr
A_0 (s) =& -\frac 1 {2\sqrt{\pi}} \frac{\Gamma (s+1/2)}{\Gamma (s+1)}
\zn s\right),\cr
A_i (s) =& -\frac 1 {\Gamma (s)} \zn s+i/ 2 \right) 
\sum_{a=0}^{2i} x_{i,a} \frac{\Gamma \left( s+(i+a)/2\right)} 
{\Gamma \left((i+a)/2 \right)}.\cr} 
\eql{12}$$
Due to the different $A_0 (s)$ and $A_i (s)$ the heat-kernel 
coefficients this time read
$$\eqalign{
A_{n/2}^{\man} =& \frac{\Gamma\left((D-n-1)/2 \right)}{\sqrt{\pi} 
(D-n)} \rzn (D-n-1)/2\right) \cr
&\hspace{*}-\frac{\Gamma\left((D-n+1)/2 \right)}{\sqrt{\pi} (D-n)}
\rzn (D-n)/2 \right) \cr
&\hspace{***}-\sum_{i=1}^{n-1} \rzn (D-n+i)/2 \right) 
\sum_{a=0}^{2i} x_{i,a} \frac{\Gamma \left( (D-n+i+a)/2 \right)}
{\Gamma \left( (i+a)/2\right) }.\cr} 
\eql{13}$$

It would be possible to rewrite the right-hand side in terms of the spinor 
heat-kernel coeffcients on the base, leading to an equation like 
(\peq{comancoeff}).

For the ball, using (\peq{resi2}) for the residues of $\zn s\right)$, 
(\peq{11}), 
all results of [\pref{DABK}] are immediately reproduced in an organised 
and rapid fashion. Let us therefore proceed to the  determinant. 

As a result of (\peq{12}) one has
$$\eqalign{
A_{-1} ' (0) =& 2\, (\ln 2 -1)\, \zn -1/2 \right) -\ze_{\can} ' \left( 
-1/2 \right),  \cr
A_0 ' (0) =& \ln 2 \,\zn 0\right) -\frac 1 2 \ze_{\can} ' (0) ,
\cr
A_i ' (0) =& -\sum_{a=0}^{2i} x_{i,a} \bigg( PP\,\, \zn i/ 2 \right)
+\gamma \rzn i/2 \right) +\cr
&\hspace{**************}\rzn i/ 2 \right) \psi \left( 
(a+i)/ 2 \right) \bigg),\cr} 
\eql{14}$$
with characteristic differences to the spectral case.
Further contributions come from
$$
\caz ' (0) = \sum_{n=0}^\infty d(n) {\caz ^{\nu}} ' (0),\quad 
\nu=n+D/2-1,\eql{15}
$$
and
$$
{\caz ^{\nu}} ' (0) = 2\ln\Gamma (\nu +1) +2\nu -2\nu\ln \nu 
-\ln (2\pi\nu ) +\sum_{n=1}^{D-1} \frac{D_n (1)}{\nu^n}. \eql{16}
$$
This follows from the results preceding eq.~(3.1) in [\pref{KandC}]. 

Formally (\peq{16}) is twice the scalar case. It  also shows that 
$D_n (1) =2\ze_R (-n) /n$ and so the contribution $\caz ' (0)$ is
known from the scalar field calculation. Adding up, everything fits nicely 
and
the final answer reads
$$\eqalign{
{\ze _{1/2}^{lo}}' (0) =& \ln 2 \big( 2\zn -1/ 2 \right) +
\zn 0\right)\big) +2\ze_{\can +1} ' (0)  \cr
&+4\ln 2 \sum_{i=1}^{D-1} \frac{\ze _R (-i) } i \rzn i/ 2 \right)
\cr
&+2\sum_{i=1}^{D-1} \rzn i/ 2 \right) \left[\frac {2\ze_R (-i)} i 
\sum_{k=1}^{i-1} \frac 1 k 
+\int_0^1 dt\,\,\frac{D_i (t) -tD_i (1)}{t(1-t^2)}\right].\cr} 
\eql{17}$$
Specialising to the ball, all results of [\pref{DABK}] are thereby 
established again in a systematic and general way. 

\section{\bf 10. Conclusion.}
Our essential results are, firstly, the expression (\peq{comancoeff}) for 
the
coexact heat-kernel coefficients on the generalised cone in terms of those 
on its base. Secondly, (\peq{genformcoeff}) gives
any ball heat-kernel coefficient, with explicitly computable polynomials,
(\peq{ppoly}), and leads to the restriction (\peq{rest}) on the general
coefficients. Thirdly, we draw attention to the formula (\peq{det}) for the
functional determinant and finally to the expression (\peq{finser}) for 
the modified $p$-form coexact \zf\ on the sphere as a finite sum of 
(conformal) 0-form functions.

To save excessive analysis, we have concentrated on the simplest generalised
cone, \ie the ball. As a continuation, more complicated bases, $\can$, can 
be
considered. In the earlier paper, [\pref{BKD}], we looked at the case when 
the
base was a sphere of non-unit radius, say $a$, calling this a monopole.
Then, in place of (\peq{nu}) for the order of the Bessel functions, we have
$$
\nu^2(l,p)={1\over a^2}\big(l+(d-1)/2\big)^2-\big((d-1)/2-p\big)^2
\big({1\over a^2}-1\big)
\eql{nu2}$$
which is not a perfect square unless $p=(d-1)/2$. We therefore expect that
a pole at $s=0$ in $\ze^\man_p(s)$ will arise even for the case of the de Rham 
Laplacian. However, unlike the scalar ($p=0$) case, it does not seem 
possible to achieve a perfect square, and hence to avoid the pole, simply by 
attaching a term to the de Rham Laplacian. This obstruction
is probably related to  the fact that there is no direct way of making the 
equation conformally invariant except for the special value $p=(d-1)/2$. 
It is possible to use Branson's operator [\pref{Branson}] but we would then 
lose contact with the body of work associated with the de Rham Laplacian.

The lack of conformal invariance can be appreciated in the structure of the
coexact \zf\ on the sphere, (\peq{finser}), which is a combination of
improved 0-form \zfs\ each of which is conformal in a different dimension.

Further possible choices for $\can$ include hyperbolic spaces and factored 
spheres for which there would be more topological and analytical excitement
involving, as they do, multiple connectedness and conical singularities. 

Regarding spectral spinors, our previous method involved fitting a 
conjectured general form for the coefficients using specific numerical
values. The present technique establishes both this general form and 
the actual expressions in one operation.
\section{\bf Acknowledgment}

We wish to thank Peter Gilkey for help regarding the algebra behind reference
[\pref{BBG}], and thanks also to Michael Bordag for interesting discussions.
KK was supported by the DFG under contract number Bo1112/4-1.

\section{\bf Appendix A.}
In this appendix we derive the form (\peq{finser}) for the coexact 
sphere \zf.
The most important property for the proof is the

\noin{\bf Lemma.}
$$\eqalign{
\sum_{m=p+1}^d \left( m-1 \atop p \right) \zb 2s; (a+1)/2 \mid {\bf 1}_m )=&
-\sum_{m=p}^{d-1} \left( m-1 \atop p-1 \right) \zb 2s; (a-1)/2 
\mid {\bf 1}_m ){}\cr
&\hspace{-2cm}+\left( d-1 \atop p \right) \zb 2s; (a-1)/2\mid{\bf 1}_d ).\cr}
\eql{l2}$$
\noin{\bf Proof.}

Rewrite the left-hand side of (\peq{l2}) using the difference relation,
[\pref{Barnesa}],
$$
\zb s; (a+1)/2\mid{\bf 1}_m )-\zb s; (a-1)/2\mid {\bf 1}_m ) = -\zb s; 
(a-1)/2\mid {\bf 1}_{m-1})
\eql{l1}
$$ 
and then combine terms using the binomial recursion (\peq{brecurs}).

Now we are set to show the following recursive property of the  
$p$-form \zf\ (\peq{spcoze})

\noin {\bf Lemma.} 
$$
\tilde \zeta_p^{S^d} (s) 
+ \tilde \zeta_{p-1}^{S^{d-2}}(s)=\left( d-1 \atop p \right)\tilde 
\ze_0^{S^d}(s)+\de_{p,1}\bigg({d-3\over2}\bigg)^{-2s}
+\de_{p,0}\bigg({d-1\over2}\bigg)^{-2s}
 \eql{l3} $$
for $0\le p\le (d-1)/2$.

\noin {\bf Proof.}

Starting from (\peq{spcoze}) and (\peq{zetzeroin}), use of (\peq{l2}) 
gives first
$$\eqalign{
\tilde\zeta_p^{S^d} (s) =&  -\sum_{m=p}^{d-1} \left( m-1 \atop p-1 
\right) \zb 2s; (d-1)/2\mid{\bf 1}_m ) +\left( d-1 \atop p \right) \zb 2s; 
(d-1)/2\mid{\bf 1}_d ){}\cr
&+\sum_{m=d-p}^d \left( m-1 \atop d-p-1\right) \zb 2s; 
(d+1)/2 \mid {\bf 1}_m )+\de_{p,0}\bigg({d-1\over2}\bigg)^{-2s}.\cr}
$$
Separate the $m=d$ term from the last sum, then
$$\eqalign{
\tilde\zeta_p^{S^d} (s) =& \left( d-1 \atop p \right)
\bigg( \zb 2s; (d+1)/2\mid {\bf 1}_d) +\zb 2s; (d-1)/2\mid {\bf 1}_d \bigg)\cr
&-\sum_{m=p}^{d-1} \left( m-1 \atop p-1\right) 
\zb 2s; (d-1)/2\mid{\bf 1}_m )+\de_{p,0}\bigg({d-1\over2}\bigg)^{-2s}\cr
&+\sum_{m=d-p}^{d-1}\left(m-1\atop d-p-1\right)\zb 2s; (d+1)/2 
\mid {\bf 1}_m ).\cr}
$$
Next, combine the $m = d-1$ terms of both sums using (\peq{l1}),
$$\eqalign{
\tilde\zeta_p^{S^d} (s) =&\left( d-1 \atop p \right)
\tilde\ze_0^{S^d}(s) -\sum_{m=p}^{d-2} \left( m-1 \atop p-1\right) 
\zb 2s; (d-1)/2\mid{\bf 1}_m )\cr
 &+\sum_{m=d-p}^{d-2}\left( m-1 \atop d-p-1\right) 
\zb 2s; (d+1)/2 \mid {\bf 1}_m )\cr
&- \left( d-2 \atop p-1 \right) \zb 2s; (d-1)/2\mid{\bf 1}_{d-2})
+\de_{p,0}\bigg({d-1\over2}\bigg)^{-2s}.\cr}
$$
Apply (\peq{l2}) once more in the second sum and then the 
binomial recursion, (\peq{brecurs}), to reach Lemma (\peq{l3}). 
Finally, applying this Lemma $p+1$ times, for $p<(d-1)/2$, 
produces the important relation ({\peq{finser}) in the main text. No doubt 
this also follows from a direct analysis of the degeneracies.

A comment on these equations is perhaps necessary. Setting $p=0$ in 
(\peq{l3}) we must recognise that $\tilde\ze_{-1}^{S^{d-2}}(s)$ equals
$\big((d-1)/2\big)^{-2s}$ for consistency. This follows 
easily from (\peq{spcoze}). In general, $\tilde\ze_{-p}^{S^{d}}(s)=
\big((d-1+2p)/2\big)^{-2s}$. 

\section{\bf Appendix B.}
In this appendix are listed some selected derivatives at zero of the 
coexact zeta 
function for absolute boundary conditions and with ${\cal M}$ the ball. 

\noindent In $d=2$,
$$\eqalign{
{\zeta_0'}^{{\cal M}}=&-{{15}\over {32}} - {{\log2}\over {12}} + \log3 
 - {{3\,\zeta'(-2)}\over 4} + {{5\,\zeta'(-1)}\over 2} + \zeta'(0)\cr  
{\zeta_1'}^{{\cal M}}=&-{1\over {16}} + {{11\,\log2}\over 6} 
- {{3\,\zeta'(-2)}\over 2} + 
   3\,\zeta'(-1) - \zeta'(0)\cr  
{\zeta_2'}^{{\cal M}}=&-{3\over {32}} - {{\log2}\over {12}} 
- {{3\,\zeta'(-2)}\over 4} + {{\zeta'(-1)}\over 2}.\cr}
$$
In $d=3$, 
$$\eqalign{
{\zeta_0'}^{{\cal M}}=&
  -{{1213}\over {4320}} +{{151\,\log2}\over {90}}
 + {{\zeta'(-3)}\over 3} + {{\zeta'(-2)}\over 2} + 
   {{13\,\zeta'(-1)}\over 6} + \zeta'(0)\cr  
{\zeta_1'}^{{\cal M}}=&{{5989}\over {10080}} - {{19\,\log2}\over {30}} 
+ \zeta'(-3) +{{\zeta'(-2)}\over 2} -{{3\,\zeta'(-1)}\over 2}-\zeta'(0)\cr  
{\zeta_2'}^{{\cal M}}=&
  -{{507}\over {1120}} + {{7\,\log2}\over {10}} 
+\zeta'(-3)-{{\zeta'(-2)}\over 2}-{{7\,\zeta'(-1)}\over 2} +2\,\zeta'(0)
\cr  
{\zeta_3'}^{{\cal M}}=&{{173}\over {30240}} + {{\log2}\over {90}} 
+{{\zeta'(-3)}\over3}-{{\zeta'(-2)}\over 2}+{{\zeta'(-1)}\over 6}
. \cr  }
$$

\noindent In $d=4$,
$$\eqalign{
{\zeta_0'}^{\cal M} =& 
  -{{25381}\over {46080}} + {{17\,\log2}\over {2880}} + \log5
 - {{5\,\zeta'(-4)}\over {64}} + {{23\,\zeta'(-3)}\over {48}}\cr
&\hspace{**************} + 
   {{47\,\zeta'(-2)}\over {32}} + {{103\,\zeta'(-1)}\over {48}} + 
   \zeta'(0)\cr
{\zeta_1'}^{\cal M} =&
  {{5803}\over {11520}} + {77\log2\over720}-\log3 - 
   {{5\,\zeta'(-4)}\over {16}} + {{19\,\zeta'(-3)}\over {12}}\cr
&\hspace{**************} +{{17\,\zeta'(-2)}\over 8}-{{25\,\zeta'(-1)}
\over{12}} - \zeta'(0)\cr
{\zeta_2'}^{\cal M} =&
  {{209}\over {2560}} - {{863\,\log2}\over {480}} - 
   {{15\,\zeta'(-4)}\over {32}} + {{15\,\zeta'(-3)}\over 8}\cr
&\hspace{**************} - 
   {{3\,\zeta'(-2)}\over {16}} - {{21\,\zeta'(-1)}\over 8} + 
   \zeta'(0)\cr
{\zeta_3'}^{\cal M} =&
  -{{2509}\over {11520}} +{77\log2\over720}- {{5\,\zeta'(-4)}\over {16}}
+{{11\,\zeta'(-3)}\over {12}}\cr
&\hspace{**************} 
    - {{7\,\zeta'(-2)}\over 8} + 
   {{19\,\zeta'(-1)}\over {12}} - \zeta'(0)\cr
{\zeta_4'}^{\cal M} =&
  {{47}\over {9216}} + {{17\,\log2}\over {2880}} - 
   {{5\,\zeta'(-4)}\over {64}} + {{7\,\zeta'(-3)}\over {48}} - 
   {{\zeta'(-2)}\over {32}} - {{\zeta'(-1)}\over {48}}.\cr}
$$
The \zfs\ on the right are Riemann \zfs.

As mentioned in the text, it is possible  to obtain the determinants in 
any dimension $d$ and for any value of $p$ without difficulty. 

\section{\bf Appendix C.}

In this appendix we outline a method for the calculation of integrals 
of the type
$$\eqalign{
\int_a^b dw\, w^{n-1}\log\Ga(w),\cr}
\eql{intneeded}
$$
needed in order to see explicitly the agreement of our results with those
of [\pref{ELV2}].

We start from the usual relation 
$$
{\pa\over \pa w}\ze(s,w)=-s\ze(s+1,w),
\eql{100}$$
valid for any modified \zf, such as $\ze^\can_p(s)$ or the Barnes \zf, 
but here applied to the Hurwitz \zf.

The differentiations are continued to give,
$$
{\pa^n\over \pa w^n}\ze(s,w)=(-1)^n s(s+1)\ldots(s+n-1)\ze(s+n,w)
=f_n(s)\ze(s+n,w).
\eql{101}$$
Now differentiate with respect to $s$
$$
{\pa^n\over \pa w^n}\ze'(s,w)=f_n'(s)\ze(s+n,w)+f_n(s)\ze'(s+n,w)
\eql{102}$$
and set $s=-n$
$$\eqalign{
{\pa^n\over \pa w^n}\ze'(-n,w)&=f_n'(-n)\ze(0,w)+f_n(-n)\ze'(0,w)\cr
&=f_n'(-n)\ze(0,w)+f_n(-n)(\log\Ga(w)-\log\sqrt{2\pi}).\cr}
\eql{103}$$
Next, multiply by $w^{n-1}$ and integrate
$$\eqalign{
f_n(-n)\int_a^b dw\, w^{n-1}&\log\Ga(w)=
\int_a^b dw\,w^{n-1}{\pa^n\over \pa w^n}\ze'(-n,w)\cr
&+{f_n(-n)\over n}(b^n-a^n)\log\sqrt{2\pi}-f_n'(-n)\int_a^b
dw\, w^{n-1}\ze(0,w).\cr}
\eql{104}$$
Define
$$
I_n(s)=\int_a^b dw\, w^{n-1}{\pa^n\over \pa w^n}\ze'(s,w)
$$and integrate by parts to get the recursion
$$
I_n(s)=w^{n-1}{\pa^{n-1}\over \pa w^{n-1}}\ze'(s,w)\bigg|_a^b-(n-1)I_{n-1}(s)
\eql{105}$$
which is trivially continued using $I_1(s)=\ze'(s,b)-\ze'(s,a)$

From (\peq{102}) we have
$$
w^{m}{\pa^{m}\over \pa w^{m}}\ze'(-n,w)=
w^{m}\big(f_{m}'(-n)\ze(-n+m,w)+f_{m}(-n)\ze'(-n+m,w)\big)
\eql{106}$$
where
$$\eqalign{
f_m(-n)=m!\comb nm,\quad f'_m(-n)=m!\comb nm \sum_{k=1}^m{1\over m-n-k}
.\cr}
\eql{anja}$$
Using the recursion (\peq{105}) $n$ times, and also (\peq{anja}), the 
following result is found for the integral (\peq{intneeded}),
$$\eqalign{
\int_a^b dw\, & w^{n-1}\log\Ga(w)= 
\bigg\{{1\over 2n} \bigg[
\log(2\pi) +\sum_{k=1}^n{1\over k}
\bigg] w^n -{1\over n+1} w^{n+1} \sum_{k=1}^n{1\over k}
\cr
&\sum_{l=1}^n {(-1)^{l-1} \over l} \comb{n-1}{l-1} w^{n-l}
\bigg[\zeta' (-l,w)-\zeta(-l,w)\sum_{k=1}^{n-l}
{1\over k+l} \bigg] \bigg\}\bigg|_a^b.\cr}
\eql{intfinal}
$$
The simplest case, when $n=1$, is given in [\pref{EandR}] and
an integral related to (\peq{intfinal}) for $n=2$ can be found in 
Nash and O'Connor, [\pref{NandOC}] eqn.(C.30).

If one of the limits in (\peq{intneeded}) is zero, the last sum in
(\peq{intfinal}) has to be interpreted as the limit as $w\to 0$ and,
in this case, we use
$$\eqalign{
\zeta (s,w) =w^{-s} +\zeta (s,1+w)\cr}
\eql{zetarec} 
$$ to see that
$$
\lim_{w \to 0} w^{n-l}\bigg[ \zeta' (-l,w) -\zeta (-l,w) 
\sum_{k=1}^{n-l} {1 \over k+l} \bigg] =0 
$$ 
for $l<n$, whereas 
$$
\lim_{w\to 0} \zeta' (-n,w) = \zeta ' (-n) 
$$
for $n\geq 1$.

Let us now give some examples of the general formula (\peq{intfinal}).
To compare with $\ze_0'(0)$ in $d=3$ of [\pref{ELV2}], we need
$$
\int_0^1 dw\,(w-1)\log\Ga(w),
$$
which is obtained by setting $n=1$ and $n=2$. 
For $n=1$, we get the well known formula ([\pref{Erdelyi}], 1.9.1 (18))
$$
\int_0^1dw\,\log\Ga(w)={1\over2}\log(2\pi)=-\ze'(0)
\eql{107}$$
which is the essential part of Raabe's formula.

For $n=2$
$$
\int_0^1 dw\,w\log\Ga(w)=-{1\over12}-{1\over2}\ze'(0)+\ze'(-1).
\eql{108}$$
The integral term in [\pref{ELV2}] is therefore
$$
2\int_0^1 dw\,(w-1)\log\Ga(w)= -{1\over6}+2\ze'(-1)+\ze'(0).
$$
In other dimensions, generally $a$ and $b$ take integer or half-integer 
values. Using (\peq{zetarec}) the integrals can always be expressed 
in terms of derivatives of the Riemann \zf. We meet some  
known integrals such as([\pref{Voros,EandR}]) 
$$
\int_1^{3/2}dw\, \log\Ga(w)=-{3\over8}-{13\over24}\log2-{3\over2}\ze'(-1)-
{1\over2}\ze'(0)
\eql{109}$$
and some lesser known ones, \eg
$$
\int_0^{3/2}dw\,w\log\Ga(w)=-{1\over2}-{11\over 16}\log2+{7\over 8}\ze'(-2)
-{3\over4}\ze'(-1)-{9\over8}\ze'(0)
\eql{110}$$
generalisations of which can be found using (\peq{intfinal}). 

Finally we point out that a generalisation of Raabe's formula to the
multiple $\Ga$-function is derived by Barnes [\pref{Barnesa}] using the
same ingredients as in the above.

\section{\bf Appendix D.}
The identity (\peq{siden1}) will now be proved.
In terms of the spinor base \zf\ it reads
$$
\rzn 1/2\right) -\rzn 1\right) -2\sum_{i=1}^{d-2} \rzn 1+i/2\right)
D_i' (1) =0   .
$$
Using knowledge of the residues in terms of the Bernoulli functions, 
[\pref{Barnesa}], one has
$$\eqalign{
0=\frac{B_{d-1}^{(d)} \left((d-1)/ 2\right)}{(d-1)!} 
&+\frac{B_{d-2}^{(d)} \left((d-1)/ 2 \right)}{(d-2)!}\cr
& +2\sum_{i=1}^{d-2} (-1)^i D_i' (1)\, \frac{B_{d-2-i}^{(d)} 
\left((d-1)/2\right)}{(1+i)! (d-2-i)!},\cr}
\eql{siden2}$$
which is a standard recursion formula for the generalized Bernoulli 
functions as will now be shown.

We first note that $D_i'(1)=(-1)^i/2$. This value is obtained by comparing 
the 
small $z$ approximation of the Bessel function $I_\nu(\nu z)$ with Olver's
asymptotic form. Thus, working always to order $z^2$,
$$
\ln I_\nu(\nu z)\sim \nu\ln\big(\nu z/2\big)-\ln\Ga(1+\nu)+\ln\bigg(
1+{1\over4(\nu+1)}(\nu z)^2\bigg).
$$
Expanding the log,
$$
\ln\bigg(1+{1\over4(\nu+1)}(\nu z)^2\bigg)\sim{1\over4} z^2{\nu\over1+1/\nu}
={1\over4}z^2\nu\sum_{n=0}^\infty(-1)^n \nu^{-n}
$$
and also
$$
\ln\Ga(1+\nu)\sim \nu\ln \nu-\nu+{1\over2}\ln2\pi\nu-\sum_{k=1}^\infty
{\ze_R(-k)\over k}\nu^{-k}.
$$ 
The Bessel asymptotics are 
$$\eqalign{
\ln I_\nu(\nu z)&\!\sim\! -{1\over2}\ln2\pi\nu\!-\!{1\over4}\ln(1+z^2)\!
+\!\nu
\bigg(\!\sqrt{1+z^2}\!+\!\ln\big({z\over1+\sqrt{1+z^2}}\big)\!\bigg)\!+\!
\sum_{n=1}^\infty\!{D_n(t)\over\nu^n}\cr
&\sim-{1\over2}\ln2\pi\nu-{1\over4}z^2+\nu\big(1+{1\over4}z^2+\ln(z/2)\big)
+\sum_{n=1}^\infty{D_n(1)-z^2D_n'(1)/2\over\nu^n}.\cr}
$$

Comparing, we obtain the quoted value of $D_n'(1)$, and also that of 
$D_n(1)$ 
met in our previous work. The higher derivatives can clearly be found by 
carrying the expansion further. 

Substituting these values of $D_n'(1)$ into (\peq{siden2}) yields the 
condition
$$
\sum_{j=0}^{d-1}\comb{d-1}j B^{(d)}_{d-1-j}\big((d-1)/2)\big)=0
\eql{cond}
$$
which is a special case of the recursion relation ([\pref{norlund2}] 
eqn.(11) p161)
$$
\sum_{j=0}^\nu\comb\nu j\om^jB^{(n)}_{\nu-j}\big(x\mid\bom\big)=
\om\nu B^{(n-1)}_{\nu-1}\big(x\mid\bom'\big),\quad \bom=(\bom',\om)
\eql{nrec}$$
involving the general Bernoulli function. Setting $\bom={\bf 1}$,
$\nu=d-1$ and $n=d$, this becomes
$$\eqalign{
\sum_{j=0}^{d-1}\comb{d-1}jB^{(d)}_{d-1-j}(x)&=(d-1)B^{(d-1)}_{d-2}(x)\cr
&=(d-1)(x-1)(x-2)\ldots(x-d+2).\cr}
\eql{srec}$$

If $d$ is odd, $(d-1)/2$ is integral and, since $1<(d-1)/2\le d-2$ for
$d>1$, the right-hand side of (\peq{srec}) vanishes for all odd $d$
when $x=(d-1)/2$, as required to show (\peq{cond}).

For even $d$, $=2q$, (\peq{srec}) allows one to give the general form
of the coefficient
$$
A^{(S)}_{(D-2)/2}(D)=(-1)^q\big(q-{1\over2}\big)\bigg(\big(q-{3\over2}\big)
\ldots {3\over2}\bigg)^2,\quad D=2q+1.
\eql{gc}$$

A similar result should hold for the monopole. Put $\om=a$ in 
(\peq{nrec}).

\section{\bf References}
\vskip 5truept
\begin{putreferences}
\ref{APS}{Atiyah,M.F., V.K.Patodi and I.M.Singer: Spectral asymmetry and 
Riemannian geometry \mpcps{77}{75}{43}.}
\ref{AandT}{Awada,M.A. and D.J.Toms: Induced gravitational and gauge-field 
actions from quantised matter fields in non-abelian Kaluza-Klein thory 
\np{245}{84}{161}.}
\ref{BandI}{Baacke,J. and Y.Igarishi: Casimir energy of confined massive 
quarks \prD{27}{83}{460}.}
\ref{Barnesa}{Barnes,E.W.: On the Theory of the multiple Gamma function 
{\it Trans. Camb. Phil. Soc.} {\bf 19} (1903) 374.}
\ref{Barnesb}{Barnes,E.W.: On the asymptotic expansion of integral 
functions of multiple linear sequence, {\it Trans. Camb. Phil. 
Soc.} {\bf 19} (1903) 426.}
\ref{Barv}{Barvinsky,A.O. Yu.A.Kamenshchik and I.P.Karmazin: One-loop 
quantum cosmology \aop {219}{92}{201}.}
\ref{BandM}{Beers,B.L. and Millman, R.S. :The spectra of the 
Laplace-Beltrami
operator on compact, semisimple Lie groups. \ajm{99}{1975}{801-807}.}
\ref{BandH}{Bender,C.M. and P.Hays: Zero point energy of fields in a 
confined volume \prD{14}{76}{2622}.}
\ref{BBG}{Bla\v zi\' k,N., Bokan,N. and Gilkey,P.B.: Spectral geometry of the 
form valued Laplacian for manifolds with boundary \ijpam{23}{92}{103-120}}
\ref{BEK}{Bordag,M., E.Elizalde and K.Kirsten: { Heat kernel 
coefficients of the Laplace operator on the D-dimensional ball}, 
\jmp{37}{96}{895}.}
\ref{BGKE}{Bordag,M., B.Geyer, K.Kirsten and E.Elizalde,: { Zeta function
determinant of the Laplace operator on the D-dimensional ball}, 
\cmp{179}{96}{215}.}
\ref{BKD}{Bordag,M., K.Kirsten,K. and Dowker,J.S.: Heat kernels and
functional determinants on the generalized cone \cmp{}{96}{}in the press.}
\ref{Branson}{Branson,T.P.: Conformally covariant equations on differential
forms \cpde{7}{82}{393-431}.}
\ref{BandG2}{Branson,T.P. and P.B.Gilkey {\it Comm. Partial Diff. Eqns.}
{\bf 15} (1990) 245.}
\ref{BGV}{Branson,T.P., P.B.Gilkey and D.V.Vassilevich {\it The Asymptotics
of the Laplacian on a manifold with boundary} II, hep-th/9504029.}
\ref{BCZ1}{Bytsenko,A.A, Cognola,G. and Zerbini, S. : Quantum fields in
hyperbolic space-times with finite spatial volume, hep-th/9605209.}
\ref{BCZ2}{Bytsenko,A.A, Cognola,G. and Zerbini, S. : Determinant of 
Laplacian on a non-compact 3-dimensional hyperbolic manifold with finite
volume, hep-th /9608089.}
\ref{CandH2}{Camporesi,R. and Higuchi, A.: Plancherel measure for $p$-forms
in real hyperbolic space, \jgp{15}{94}{57-94}.} 
\ref{CandH}{Camporesi,R. and A.Higuchi {\it On the eigenfunctions of the 
Dirac operator on spheres and real hyperbolic spaces}, gr-qc/9505009.}
\ref{ChandD}{Chang, Peter and J.S.Dowker :Vacuum energy on orbifold factors
of spheres, \np{395}{93}{407}.}
\ref{cheeg1}{Cheeger, J.: Spectral Geometry of Singular Riemannian Spaces.
\jdg {18}{83}{575}.}
\ref{cheeg2}{Cheeger,J.: Hodge theory of complex cones {\it Ast\'erisque} 
{\bf 101-102}(1983) 118-134}
\ref{Chou}{Chou,A.W.: The Dirac operator on spaces with conical 
singularities and positive scalar curvature, \tams{289}{85}{1-40}.}
\ref{CandT}{Copeland,E. and Toms,D.J.: Quantized antisymmetric 
tensor fields and self-consistent dimensional reduction 
in higher-dimensional spacetimes, \break\np{255}{85}{201}}
\ref{DandH}{D'Eath,P.D. and J.J.Halliwell: Fermions in quantum cosmology 
\prD{35}{87}{1100}.}
\ref{cheeg3}{Cheeger,J.:Analytic torsion and the heat equation. \aom{109}
{79}{259-322}.}
\ref{DandE}{D'Eath,P.D. and G.V.M.Esposito: Local boundary conditions for 
Dirac operator and one-loop quantum cosmology \prD{43}{91}{3234}.}
\ref{DandE2}{D'Eath,P.D. and G.V.M.Esposito: Spectral boundary conditions 
in one-loop quantum cosmology \prD{44}{91}{1713}.}
\ref{Dow1}{Dowker,J.S.: Effective action on spherical domains, \cmp{162}{94}
{633}.}
\ref{Dow8}{Dowker,J.S. {\it Robin conditions on the Euclidean ball} 
MUTP/95/7; hep-th\break/9506042. {\it Class. Quant.Grav.} to be published.}
\ref{Dow9}{Dowker,J.S. {\it Oddball determinants} MUTP/95/12; 
hep-th/9507096.}
\ref{Dow10}{Dowker,J.S. {\it Spin on the 4-ball}, 
hep-th/9508082, {\it Phys. Lett. B}, to be published.}
\ref{DandA2}{Dowker,J.S. and J.S.Apps, {\it Functional determinants on 
certain domains}. To appear in the Proceedings of the 6th Moscow Quantum 
Gravity Seminar held in Moscow, June 1995; hep-th/9506204.}
\ref{DABK}{Dowker,J.S., Apps,J.S., Bordag,M. and Kirsten,K.: Spectral 
invariants for the Dirac equation with various boundary conditions 
{\it Class. Quant.Grav.} to be published, hep-th/9511060.}
\ref{EandR}{E.Elizalde and A.Romeo : An integral involving the
generalized zeta function, {\it International J. of Math. and 
Phys.} {\bf13} (1994) 453.}
\ref{ELV2}{Elizalde, E., Lygren, M. and Vassilevich, D.V. : Zeta function 
for the laplace operator acting on forms in a ball with gauge boundary 
conditions. hep-th/9605026}
\ref{ELV1}{Elizalde, E., Lygren, M. and Vassilevich, D.V. : Antisymmetric
tensor fields on spheres: functional determinants and non-local
counterterms, \jmp{}{96}{} to appear. hep-th/ 9602113}
\ref{Kam2}{Esposito,G., A.Y.Kamenshchik, I.V.Mishakov and G.Pollifrone: 
Gravitons in one-loop quantum cosmology \prD{50}{94}{6329}; 
\prD{52}{95}{3457}.}
\ref{Erdelyi}{A.Erdelyi,W.Magnus,F.Oberhettinger and F.G.Tricomi {\it
Higher Transcendental Functions} Vol.I McGraw-Hill, New York, 1953.}
\ref{Esposito}{Esposito,G.: { Quantum Gravity, Quantum Cosmology and 
Lorentzian Geometries}, Lecture Notes in Physics, Monographs, Vol. m12, 
Springer-Verlag, Berlin 1994.}
\ref{Esposito2}{Esposito,G. {\it Nonlocal properties in Euclidean Quantum
Gravity}. To appear in Proceedings of 3rd. Workshop on Quantum Field Theory
under the Influence of External Conditions, Leipzig, September 1995; 
gr-qc/9508056.}
\ref{EKMP}{Esposito G, Kamenshchik Yu A, Mishakov I V and Pollifrone G.:
One-loop Amplitudes in Euclidean quantum gravity.
\prd {52}{96}{3457}.}
\ref{ETP}{Esposito,G., H.A.Morales-T\'ecotl and L.O.Pimentel {\it Essential
self-adjointness in one-loop quantum cosmology}, gr-qc/9510020.}
\ref{FORW}{Forgacs,P., L.O'Raifeartaigh and A.Wipf: Scattering theory, U(1) 
anomaly and index theorems for compact and non-compact manifolds 
\np{293}{87}{559}.}
\ref{GandM}{Gallot S. and Meyer,D. : Op\'erateur de coubure et Laplacian
des formes diff\'eren-\break tielles d'une vari\'et\'e riemannienne 
\jmpa{54}{1975}
{289}.}
\ref{Gilkey1}{Gilkey, P.B, Invariance theory, the heat equation and the
Atiyah-Singer index theorem, 2nd. Edn., CTC Press, Boca Raton 1995.}
\ref{Gilkey2}{Gilkey,P.B.:On the index of geometric operators for 
Riemannian manifolds with boundary \aim{102}{93}{129}.}
\ref{Gilkey3}{Gilkey,P.B.: The boundary integrand in the formula for the 
signature and Euler characteristic of a manifold with boundary 
\aim{15}{75}{334}.}
\ref{Grubb}{Grubb,G. {\it Comm. Partial Diff. Eqns.} {\bf 17} (1992) 
2031.}
\ref{GandS1}{Grubb,G. and R.T.Seeley \cras{317}{1993}{1124}; \invm{121}{95}
{481}.}
\ref{GandS}{G\"unther,P. and Schimming,R.:Curvature and spectrum of compact
Riemannian manifolds, \jdg{12}{77}{599-618}.}
\ref{IandT}{Ikeda,A. and Taniguchi,Y.:Spectra and eigenforms of the 
Laplacian
on $S^n$ and $P^n(C)$. \ojm{15}{1978}{515-546}.}
\ref{IandK}{Iwasaki,I. and Katase,K. :On the spectra of Laplace operator
on $\La^*(S^n)$ \pja{55}{79}{141}.}
\ref{JandK}{Jaroszewicz,T. and P.S.Kurzepa: Polyakov spin factors and 
Laplacians on homogeneous spaces \aop{213}{92}{135}.}
\ref{Kam}{Kamenshchik,Yu.A. and I.V.Mishakov: Fermions in one-loop quantum 
cosmology \prD{47}{93}{1380}.}
\ref{KandM}{Kamenshchik,Yu.A. and I.V.Mishakov: Zeta function technique for
quantum cosmology {\it Int. J. Mod. Phys.} {\bf A7} (1992) 3265.}
\ref{KandC}{Kirsten,K. and Cognola.G,: { Heat-kernel coefficients and 
functional determinants for higher spin fields on the ball} \cqg{13}{96}
{633-644}.}
\ref{Levitin}{Levitin,M.: { Dirichlet and Neumann invariants for Euclidean
balls}, {\it Diff. Geom. and its Appl.}, to be published.}
\ref{Luck}{Luckock,H.C.: Mixed boundary conditions in quantum field theory 
\jmp{32}{91}{1755}.}
\ref{MandL}{Luckock,H.C. and Moss,I.G,: The quantum geometry of random 
surfaces and spinning strings \cqg{6}{89}{1993}.}
\ref{Ma}{Ma,Z.Q.: Axial anomaly and index theorem for a two-dimensional 
disc 
with boundary \jpa{19}{86}{L317}.}
\ref{Mcav}{McAvity,D.M.: Heat-kernel asymptotics for mixed boundary 
conditions \cqg{9}{92}{1983}.}
\ref{MandV}{Marachevsky,V.N. and D.V.Vassilevich {\it Diffeomorphism
invariant eigenvalue \break problem for metric perturbations in a bounded 
region}, SPbU-IP-95, \break gr-qc/9509051.}
\ref{Milton}{Milton,K.A.: Zero point energy of confined fermions 
\prD{22}{80}{1444}.}
\ref{MandS}{Mishchenko,A.V. and Yu.A.Sitenko: Spectral boundary conditions 
and index theorem for two-dimensional manifolds with boundary 
\aop{218}{92}{199}.}
\ref{Moss}{Moss,I.G.: Boundary terms in the heat-kernel expansion 
\cqg{6}{89}{759}.}
\ref{MandP}{Moss,I.G. and S.J.Poletti: Conformal anomaly on an Einstein space 
with boundary \pl{B333}{94}{326}.}
\ref{MandP2}{Moss,I.G. and S.J.Poletti \np{341}{90}{155}.}
\ref{NandOC}{Nash, C. and O'Connor,D.J.: Determinants of Laplacians, the 
Ray-Singer torsion on lens spaces and the Riemann zeta function 
\jmp{36}{95}{1462}.}
\ref{NandS}{Niemi,A.J. and G.W.Semenoff: Index theorem on open infinite 
manifolds \np {269}{86}{131}.}
\ref{NandT}{Ninomiya,M. and C.I.Tan: Axial anomaly and index thorem for 
manifolds with boundary \np{245}{85}{199}.}
\ref{norlund2}{N\"orlund~N. E.:M\'emoire sur les polynomes de Bernoulli.
\am {4}{21} {1922}.}
\ref{Poletti}{Poletti,S.J. \pl{B249}{90}{355}.}
\ref{RandT}{Russell,I.H. and Toms D.J.: Vacuum energy for massive forms 
in $R^m\times S^N$, \cqg{4}{86}{1357}.}
\ref{RandS}{R\"omer,H. and P.B.Schroer \pl{21}{77}{182}.}
\ref{Trautman}{Trautman,A.: Spinors and Dirac operators on hypersurfaces 
\jmp{33}{92}{4011}.}
\ref{Vass}{Vassilevich,D.V.{Vector fields on a disk with mixed 
boundary conditions} gr-qc /9404052.}
\ref{Voros}{Voros,A.:
Spectral functions, special functions and the Selberg zeta function.
\cmp{110}{87}439.}

\end{putreferences}

\bye